\documentclass[10pt,twocolumn]{wlscirep}
\usepackage{upgreek}
\usepackage{soul}

\usepackage{setspace}

\usepackage[pagewise]{lineno}

\title{\LARGE Broadband Mie-driven random quasi-phase-matching}

\author[1,*]{Romolo Savo}
\author[1]{Andrea Morandi}
\author[1]{Jolanda S. M\"uller}
\author[1]{Fabian Kaufmann}
\author[1]{Flavia Timpu}
\author[1]{Marc Reig Escalé}
\author[2,3]{Michele Zanini}
\author[2]{Lucio Isa}
\author[1]{Rachel Grange}
\affil[1]{Optical Nanomaterial Group, Institute for Quantum Electronics,Department of Physics, ETH Zurich, Auguste-Piccard-Hof 1, 8093 Zurich,Switzerland}
\affil[2]{Laboratory for Soft Materials and Interfaces, Department of Materials, ETH Zurich, Vladimir-Prelog- Weg 5,8093 Zurich, Switzerland}
\affil[3]{FenX AG, Vladimir-Prelog Weg 5, 8093 Zurich}

\affil[*]{savor@phys.ethz.ch}

\begin{abstract}
\bf
High-quality crystals without inversion symmetry are the conventional platform to achieve optical frequency conversion via three wave-mixing.
In bulk crystals, efficient wave-mixing relies on phase-matching configurations, while at the micro- and nano-scale it requires resonant mechanisms that enhance the nonlinear light-matter interaction. These strategies commonly result in wavelength-specific performances and narrowband applications. 
Disordered photonic materials, made up of a random assembly of optical nonlinear crystals, enable a broadband tunability in the random quasi-phase-matching (RQPM) regime and do not require  high-quality materials. 
Here, we combine resonances and disorder by implementing RQPM in Mie-resonant spheres of a few microns realized by the bottom-up assembly of barium titanate nano-crystals. The measured second harmonic generation (SHG) reveals a combination of broadband and resonant wave mixing, in which Mie resonances drive and enhance the SHG, while the disorder keeps the phase-matching conditions relaxed.
This new phase-matching regime can be described by a random walk in the SHG complex plane whose step lengths depend on the local field enhancement within the micro-sphere.
Our nano-crystals assemblies provide new opportunities for tailored phase-matching at the micro-scale, beyond the coherence length of the bulk crystal. They can be adapted to achieve frequency conversion from the near-ultraviolet to the infrared ranges, they are low-cost and scalable to large surface areas. 

\end{abstract}

\begin{document}


\maketitle

\noindent 
\newpage


Nonlinear optical processes of the second order -  mediated by a $\chi^{(2)}$ susceptibility - are the common means to obtain coherent light at wavelengths not available with laser sources~\cite{boyd2008nonlinear} and have become a reliable way to generate photon quantum states~\cite{KwiatZeilinger1995_PRL_PhotonPairs}. Applications are relevant in spectroscopy~\cite{Shen:1996vl,Werner2015OL}, bio-imaging~\cite{campagnola2003second}, ultrafast optics~\cite{trebino2012frequency} and quantum photonics~\cite{caspani2017integrated}, making these nonlinear processes key to the development of near future photonic technology. %
Non-centrosymmetric crystals, which lack inversion symmetry, are among the most attractive materials with second-order nonlinearity, thanks to high $\chi^{(2)}$ coefficients, wide transparency windows (visible to near-infrared) and high damage thresholds~\cite{gunter2012nonlinear}. 
As known, their use is bounded by optical dispersion, which imposes strict phase-matching conditions to achieve optimal nonlinear conversion. Many methods have been developed for phase-matching control, as  phase matching in birefringent crystals~\cite{boyd2008nonlinear}, quasi-phase matching~\cite{fejer1992QPM}, modal~\cite{moutzouris2003modalPM} and cyclic~\cite{lin2013cyclicPM} phase-matching. Strategies to enhance the nonlinear process at scales smaller than the coherence length of the crystal are also available, as cavity coupling~\cite{rivoire2011cavity,lin2016cavity}, vanishing-permittivity materials~\cite{suchowski2013phase}, plasmonic and dielectric nano-resonators~\cite{pu2010coreshell}.
In all cases, the optimization of the nonlinear processes is based on an underlying resonant mechanism, providing wavelength-specific performances that hamper the use of $\chi^{(2)}$ crystalline devices for broadband, widely tunable applications. 

Increasing the complexity of the nonlinear crystal structure leads to relaxed phase-matching conditions and to largely improved performances in broadband nonlinear mixing~\cite{suchowski2010broadband}. In the case of a fully disordered crystal, characterized by polydispersed and randomly arranged non-centrosymmetric crystalline domains, the amplitudes and phases of the mixed waves get randomized, so that interference terms average out and the nonlinear signal accumulates as the sum of the intensities generated from each domain. This three-wave mixing regime is known as random quasi-phase matching~\cite{baudrier2004random, skipetrov2004RQPM, vidal2006generation}(RQPM) and so far it has been implemented in disordered polycrystals with micron-sized domains(10-100 $\upmu$m)~\cite{baudrier2004random,bravo2010optical,ru2017OPOrandom}.
Its distinctive feature is the linear scaling of the nonlinear generated power with the number of domains, over distances larger than the coherence length and without any geometric constraints~\cite{baudrier2004random}. The disordered distribution of the crystalline domains also enables a flat broadband tunability of the wave-mixing, due to the absence of resonant optimization mechanisms, as demonstrated in naturally grown crystals with a two-dimensional disorder~\cite{fischer2006broadband, molina2008NLphotglass}.
\begin{figure*}[ht!]
   \centering
   \includegraphics[scale=1]{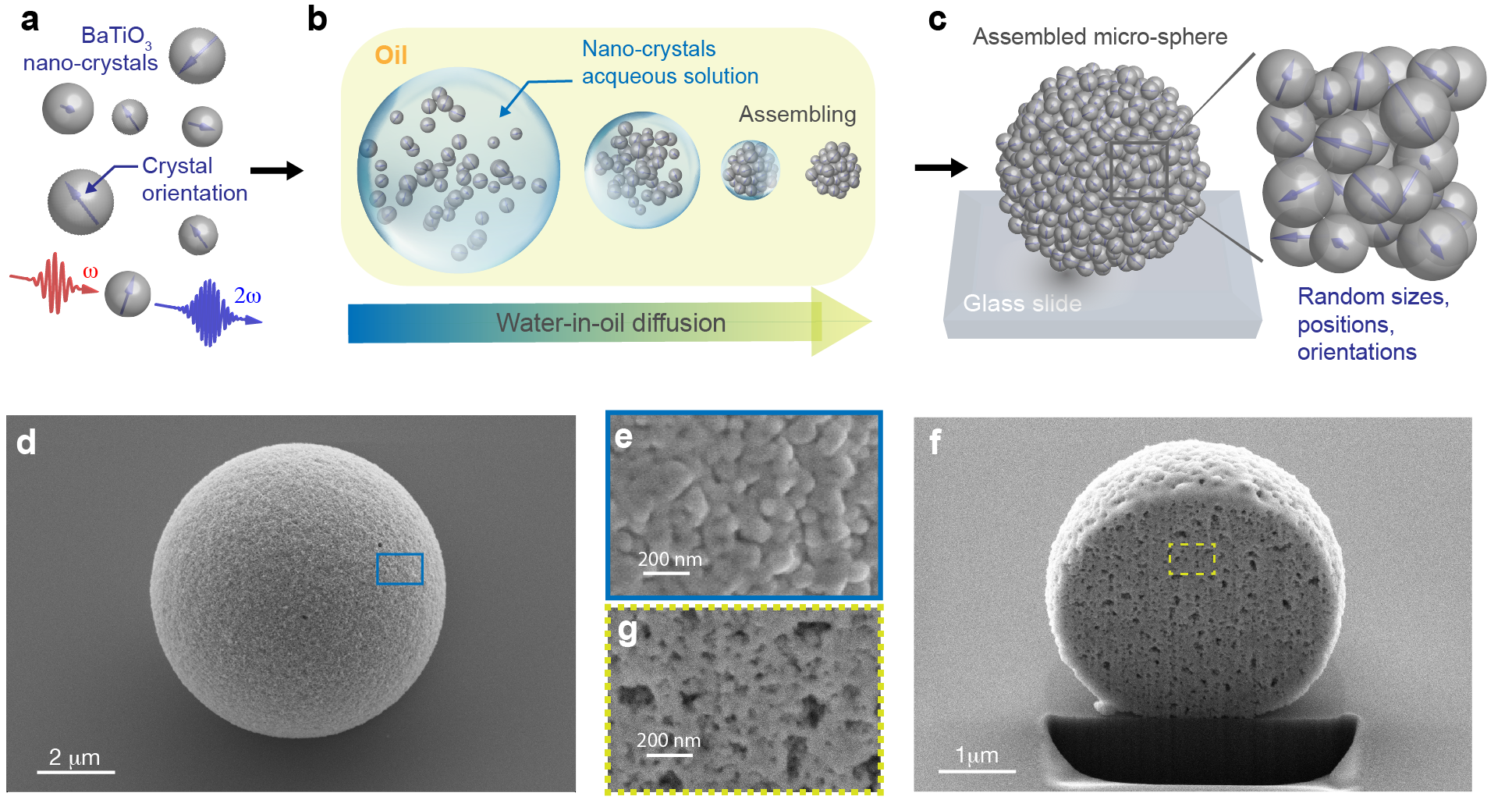}
   \caption{\textbf{Bottom-up assembly of BaTiO$_3$ disordered micro-spheres.} \textbf{a}, Representation of the BaTiO$_3$ nano-crystals used in the assembly procedure. Their SHG efficiency depends on the size and orientation of the crystal, here indicated by the arrows. \textbf{b}, Sketch of the emulsion-driven assembly procedure. \textbf{c}, Representation of an assembled micro-sphere highlighting the randomness in the sizes, positions and orientations of the nano-crystals. \textbf{d}, SEM image of a BaTiO$_3$ micro-sphere assembled on a silicon substrate for better image quality. \textbf{e}, Close-up of the surface of the micro-sphere highlighting the disordered arrangement of the crystalline nano-domains. \textbf{f}, SEM image of the cross-section of a BaTiO$_3$ micro-sphere obtained by FIB. \textbf{g}, Close-up of the disordered nano-porous inner structure of the micro-sphere}
   \label{fig:figure1}
\end{figure*}

A completely unexplored aspect, which we address here, is the use of non-centrosymmetric crystalline nano-domains to realize miniaturized $\chi^{(2)}$ disordered structures with a controlled geometry. This way, the system could sustain geometric resonances that enhance the nonlinear wave-mixing, similarly to crystalline micro-~\cite{ricciardi2015frequency} and nano-resonators~\cite{kuznetsov2016optically}, simultaneously showing relaxed phase-matching conditions and a broadband tunability thanks to RQPM.  
The use of nano-domains is further motivated by the availability of  metal-oxide  nano-crystals with bulk $\chi^{(2)}$ properties~\cite{kim2013second,timpu2016second}, which allows for the bottom-up fabrication of $\chi^{(2)}$ disordered structures with an a priori control over the domain size distribution.
So far, no observation of RQPM in nano-structured disorder has been reported, although theory does not predict limitations on the possible domain size~\cite{vidal2006generation}. Attention has been dedicated to three-wave mixing in turbid crystalline nano-powders~\cite{deBoer1993SHGcorr}, but only in the multiple scattering regime~\cite{faez2009SHGdiff,makeev2003second}. Resonant enhancement of RQPM has been observed in a large optical-parametric-oscillator cavity~\cite{ru2017OPOrandom}, but never at the micro-scale.

Here, we realize three-dimensional $\chi^{(2)}$ disordered micro-spheres by bottom-up assembly of barium titanate (BaTiO$_3$) nano-crystals and demonstrate their second harmonic generation (SHG) through RQPM. This is identified by the linear scaling of the SHG power with the volume of the micro-structures, over more than three orders of magnitude, reaching sizes six times larger than the coherence length of BaTiO$_3$. Numerical modelling shows that the efficiency of this process is comparable with that of crystalline BaTiO$_3$ of the same size, with the remarkable advantage of having a monotonic growth with the structure size and relaxed phase-matching  conditions for the illumination direction and polarization.
Thanks to their homogeneity in the refractive index, assembled micro-spheres sustain high-order Mie resonances stemming from their outer geometry, which couple to the RQPM mechanism and drive the SHG. Sweeping of the pump wavelength over a 100~nm range reveals a modulated SHG, resulting from a unique combination of broadband  and resonant wave mixing. This new regime of RQPM is described accurately by a random walk in the SHG complex plane whose step lengths depend on the local field enhancement within the micro-spheres.
We have realized a new class of nonlinear $\chi^{(2)}$ resonators by bottom-up assembly of nano-crystals, providing evidence and modelling of an unexplored phase-matching regime, in which random quasi-phase matching and Mie resonances couple together. Our systems are low-cost and easily scalable to large surface areas, opening up to new designs for broadband and tunable nonlinear photonic devices based on disorder.  
\section*{Results}
\subsection*{Bottom-up assembly of $\chi^{(2)}$ disordered  BaTiO$_3$ micro-spheres }

We realized micron-sized spherical structures by emulsion-templated assembly~\cite{kim2008microspheres,vogel2015color} of colloidal barium titanate (BaTiO$_3$) nano-crystals (mean diameter 50~nm, 5\% polydispersity).
As previously empirically confirmed~\cite{VoglerNeuling_pssb2020}, the synthetic nano-crystals are characterized by a crystallographic tetragonal phase enabling bulk SHG at the nano-scale under femtosecond pulsed illumination~\cite{pu2010coreshell,timpu2016second}, see sketch in Fig~\ref{fig:figure1}a.
We developed an assembly procedure (see Methods) in which the aqueous dispersion (2wt$\%$) of BaTiO$_3$ nano-crystals is mixed with surfactant-loaded hexadecane (SPAN80 1wt$\%$ ) and emulsified by mechanical shaking to generate  polydispersed water-in-oil droplets, as sketched in Fig.~\ref{fig:figure1}b. Water evaporation through the hexadecane reduces the size of the droplets and the water-oil interfaces act as dynamic templates for the assembly of the nano-crystals into larger micro-spheres. Assemblies are finally deposited on a glass substrate,~Fig.~\ref{fig:figure1}b-c. 
Their size distribution depends on the size dispersion of the droplets and on the concentration of the nano-crystals in water. We generated micro-spheres with diameters from 0.5~$\upmu$m to 20~$\upmu$m. Assembled structures have a purely spherical geometry, Fig.~\ref{fig:figure1}d, over the whole size range (see supplementaries S2). The surface roughness is solely determined by the finite size of the nano-crystals. 
A close-up of the micro-sphere surface is shown in Fig.~\ref{fig:figure1}e, where the crystalline polydispersed nano-domains are clearly visible.  
Due to the Brownian motion and to the polydispersity, nano-crystals assemble with random orientations and positions, as sketched in Fig.~\ref{fig:figure1}c. 
The micro-sphere cross section, shown in Fig.~\ref{fig:figure1}f-g, uncovers a disordered nano-porous structure with a filling fraction of 55$\%$, measured by image analysis (see supplementaries S3).
The micro-spheres are not sintered and the nano-crystals are bound by surface forces. Remarkably, they are robust and appear free from deformation several months after the fabrication. 
%
%
%
\subsection*{Linear effective-medium-Mie behavior} 
\begin{figure}[ht!]
   \centering
   \includegraphics[scale=1]{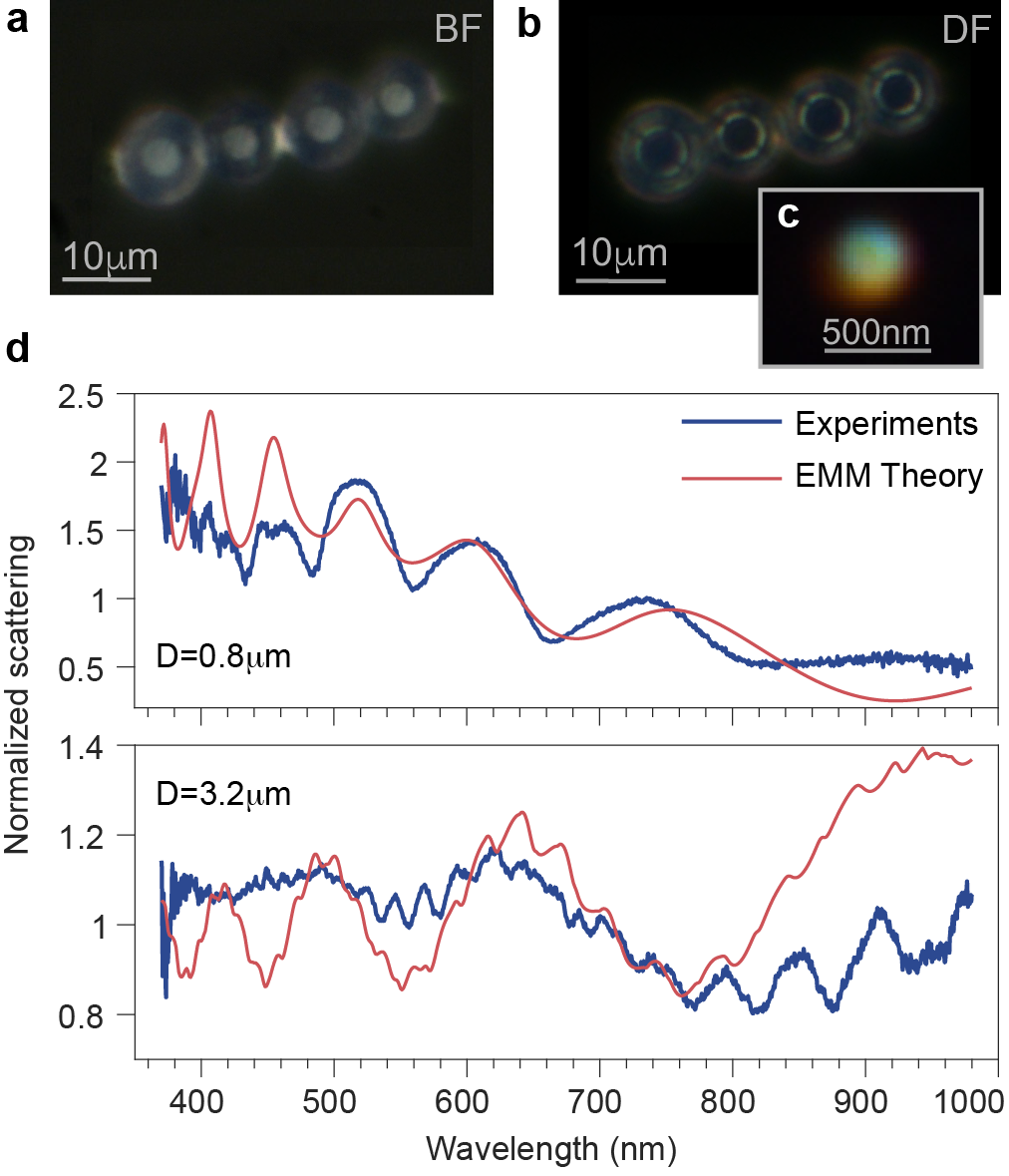}
   \caption{\textbf{Linear optical characterization of assembled micro-spheres}. \textbf{a}, Optical image of a group of micro-spheres acquired in reflection under BF illumination. The spot in the center of the micro-spheres reproduces the corresponding K\"ohler illumination. \textbf{b}, Same as in (\textbf{a}), but under DF illumination, reproducing the crossed ring of light of the source. \textbf{c}, Optical image of a sub-micron sphere acquired in reflection under DF illumination. At this scale, only a colorful spot becomes visible. \textbf{d}, Light-scattering spectra measured in DF for two micro-spheres with different diameters D. Experiments are fitted with the effective-medium-Mie (EMM) model. Best-fit curves correspond to filling fraction of 52$\%$ for D=0.8~$\upmu$m and 55$\%$ for D=3.2~$\upmu$m, in agreement with the filling fraction measured by image analysis. For both curves, the absorption coefficient was set to $k=0.003$. }
   \label{fig:figure2}
\end{figure}
\begin{figure}[ht!]
   \centering
   \includegraphics[scale=1]{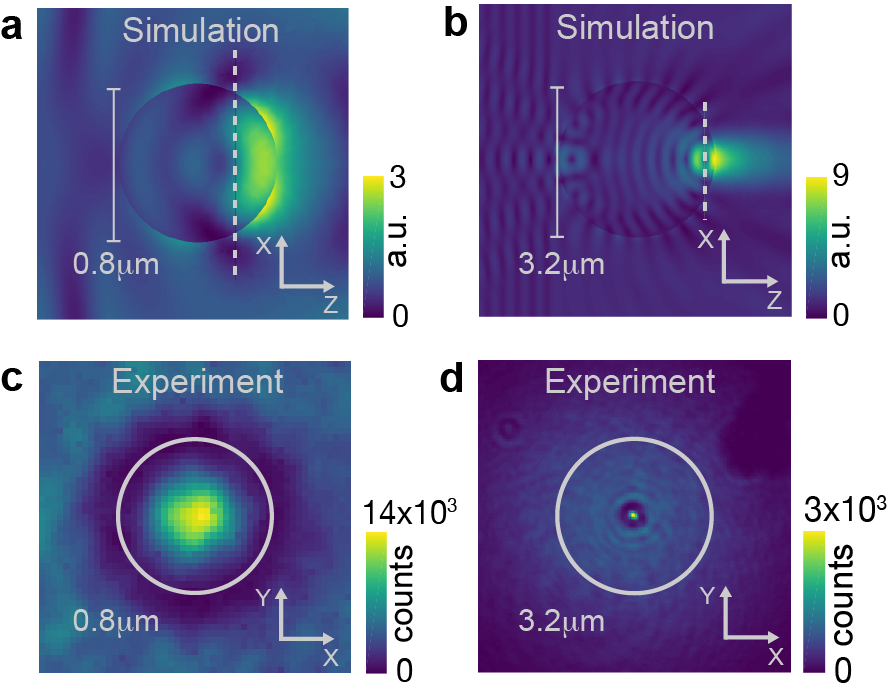}
   \caption{\textbf{Simulation and visualization of the linear quasi-normal modes of the micro-spheres}. \textbf{a}, Finite-element-simulation (FEM) in the effective-medium approximation of the linear modes excited by a plane wave (coming from the left) at 930~nm in a micro-sphere with D=0.8~$\upmu$m. The pattern shows the intensity distribution of the modes  in the x-z plane. \textbf{b}, FEM simulation as in (\textrm{a}) for a micro-sphere with D=3.2~$\upmu$m. \textbf{c}, Experimental image of the rear plane of a micro-sphere with D=0.8~$\upmu$m obtained with laser light at 930~nm. The line indicates the diameter of the micro-sphere measured by SEM. The plane of this image is orthogonal to the one simulated in (\textbf{a}), where it is indicated by the white dashed line.  \textbf{d}, Experimental image as in (\textbf{b}) for a micro-sphere with D=3.2~$\upmu$m. The plane of this image is indicated in (\textbf{b}) by the white dashed line}
   \label{fig:figure3}
\end{figure}
Linear optical properties of the assembled micro-spheres have been investigated with an upright microscope customized for spectral measurements to identify resonant scattering (see supplementaries S4). Visual inspection under bright-field (BF) and dark-field (DF) illumination provides first evidence that the nano-crystals-air composite behaves as an effective medium. Indeed, the micro-spheres appear optically homogeneous with a sufficient clearness to function as micro-lenses and to create an image of the illumination source~\cite{yang2014super}, Fig.~\ref{fig:figure1}a-c.
Spectra of the  light scattered under DF illumination are shown in Fig.~\ref{fig:figure2}d, for two different diameters. We observe distinct resonances that get spectrally denser for the larger diameter, since higher-order modes are excited, showing that the micro-spheres act as optical Mie resonators~\cite{checcucci2018titania}. 
To describe the observation and to unambiguously relate it to the spherical geometry of the assembly, we modelled the micro-spheres as homogeneous spheres with the same size and with an effective refractive index. Then, we calculated the scattering cross section with Mie theory. We refer to this description as an effective-medium-Mie (EMM) model (see Methods and supplementaries S5).
The good agreement between experiments and theory shown in Fig.~\ref{fig:figure2}d  supports the Mie-resonator description. The EMM model estimates an effective refractive index $n_\textrm{eff}\approx$1.55 and highlights the widening of the resonances due to the presence of the substrate, which we have taken into account by adding an imaginary part to $n_\textrm{eff}$ (see Methods and supplementaries S5).  %
 The more qualitative agreement obtained for the larger structure is determined by the limited field-of-view of the DF configuration ($\approx$4$\upmu$m). For micro-spheres of similar size, part of the scattered light is not collected. Moreover, the increased scattering probability within the larger micro-spheres reduces the applicability of the EMM model. 
 The quasi-normal modes~\cite{lalanne2018light} of the two micro-spheres have been computed by finite-element-methods (FEM) simulations in the effective-medium approximation, Fig.~\ref{fig:figure3}a-b.
 The pronounced mode confinement opposite to the illumination is a focusing effect known as a \textit{photonic nanojet}~\cite{chen2004photonic}.
We experimentally observed the nanojets in both micro-spheres by imaging the micro-spheres rear plane under coherent illumination with the setup sketched in Fig.~\ref{fig:figure4}a and by collecting only the pump light at 930~nm. The spatial features of the observed nanojets, Fig.~\ref{fig:figure3}c-d, match the simulated ones, providing the first observation of photonic nanojets in a composite micro-sphere up to our knowledge.
Since the nanojet can be derived by Mie theory~\cite{Geints2012_nanojet}, both the resonant and the focusing properties observed experimentally corroborate the EMM behavior of the micro-spheres.
%
\subsection*{SHG through Random Quasi-Phase-Matching}
\begin{figure*}[ht!]
   \centering
   \includegraphics[scale=1]{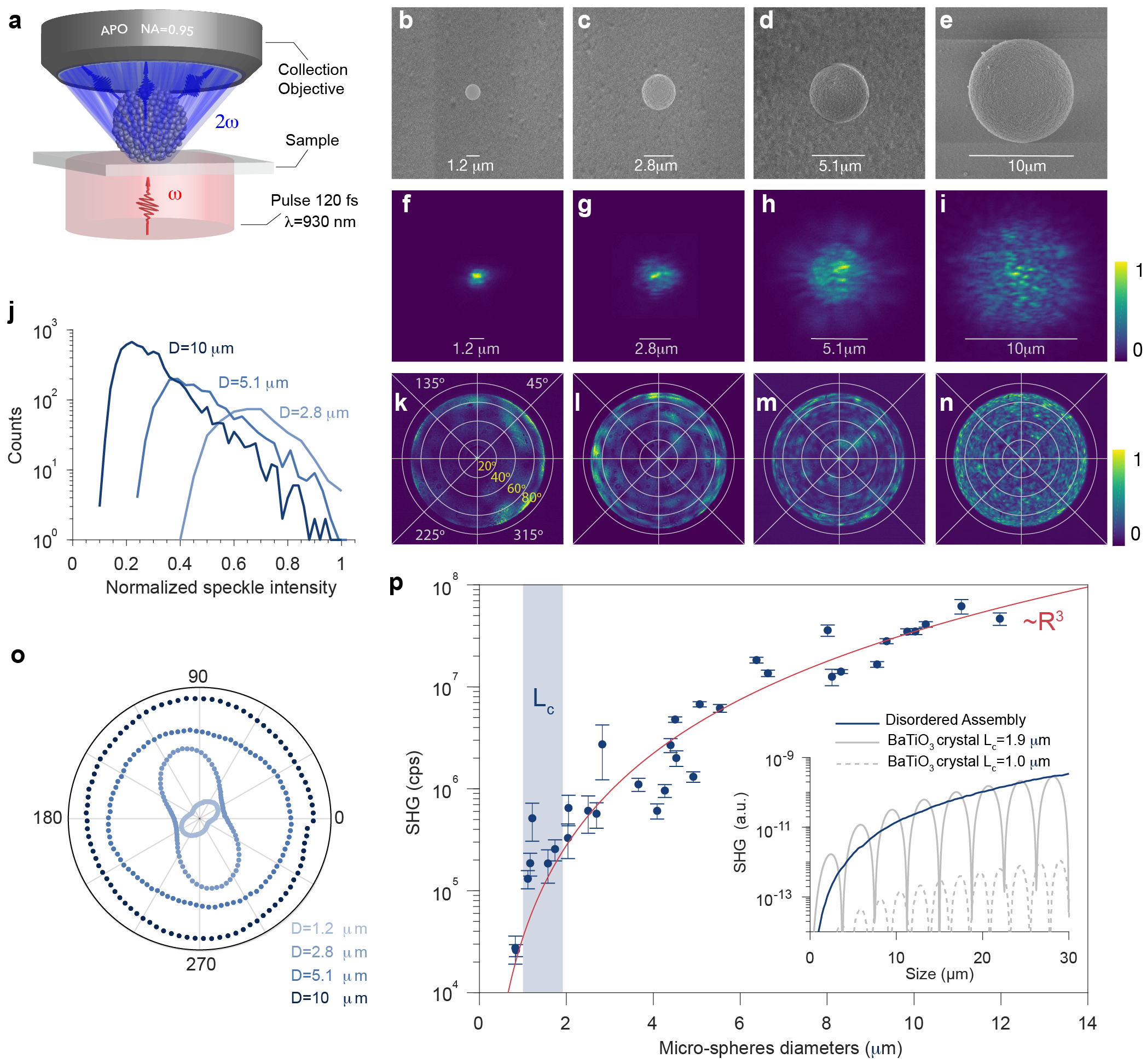}
   \caption{\textbf{SHG from the assembled micro-spheres and observation of RQPM.} \textbf{a}, Sketch of the experimental configuration. \textbf{b-e}, SEM images of four micro-spheres assembled on a glass substrate with increasing size. Diameters are indicated. \textbf{f-i}, Images of the SHG emitted from (\textbf{b-e}). Correspondence is along the column. The focal plane of the objective is placed in the central section of the structures. Reported sizes are measured from the image after calibration of the camera. \textbf{j}, Intensity distributions for the speckles in (\textbf{g-i}). The y-axis is in log-scale. \textbf{k-n}, Images of the  back-focal-plane (BFP) of the objective for the SHG shown in (\textbf{f-i}). Correspondence is along the column. \textbf{o}, SHG integrated over the entire speckle image for a variable input polarisation. Measurements are performed on the micro-spheres in (\textbf{b-e}). \textbf{p}, SHG from a set of 32 micro-spheres with increasing diameter,  integrated over the entire speckle image and averaged over the input polarization. Error bars report the range of variability when rotating the input polarization. The red line is a fit of the data with the function $y=ax^3$. The blue bar indicates the range of variability of the coherence lengths (L$_\textrm{c}$) of BaTiO$_3$  at 930~nm. The inset  shows the numerical comparison between the SHG from disordered assemblies of BaTiO$_3$ with increasing size (nano-crystals of 50~nm), and crystals of  BaTiO$_3$ of the same size, for the maximum and for a randomly chosen coherence length. All structures have a cubic shape and a uniform pump illumination. The x-axis reports the side length. The same calculations in terms of the SHG efficiency are reported in the SI. }
   \label{fig:figure4}
\end{figure*}
\begin{figure*}[h!]
   \centering
   \includegraphics[scale=1]{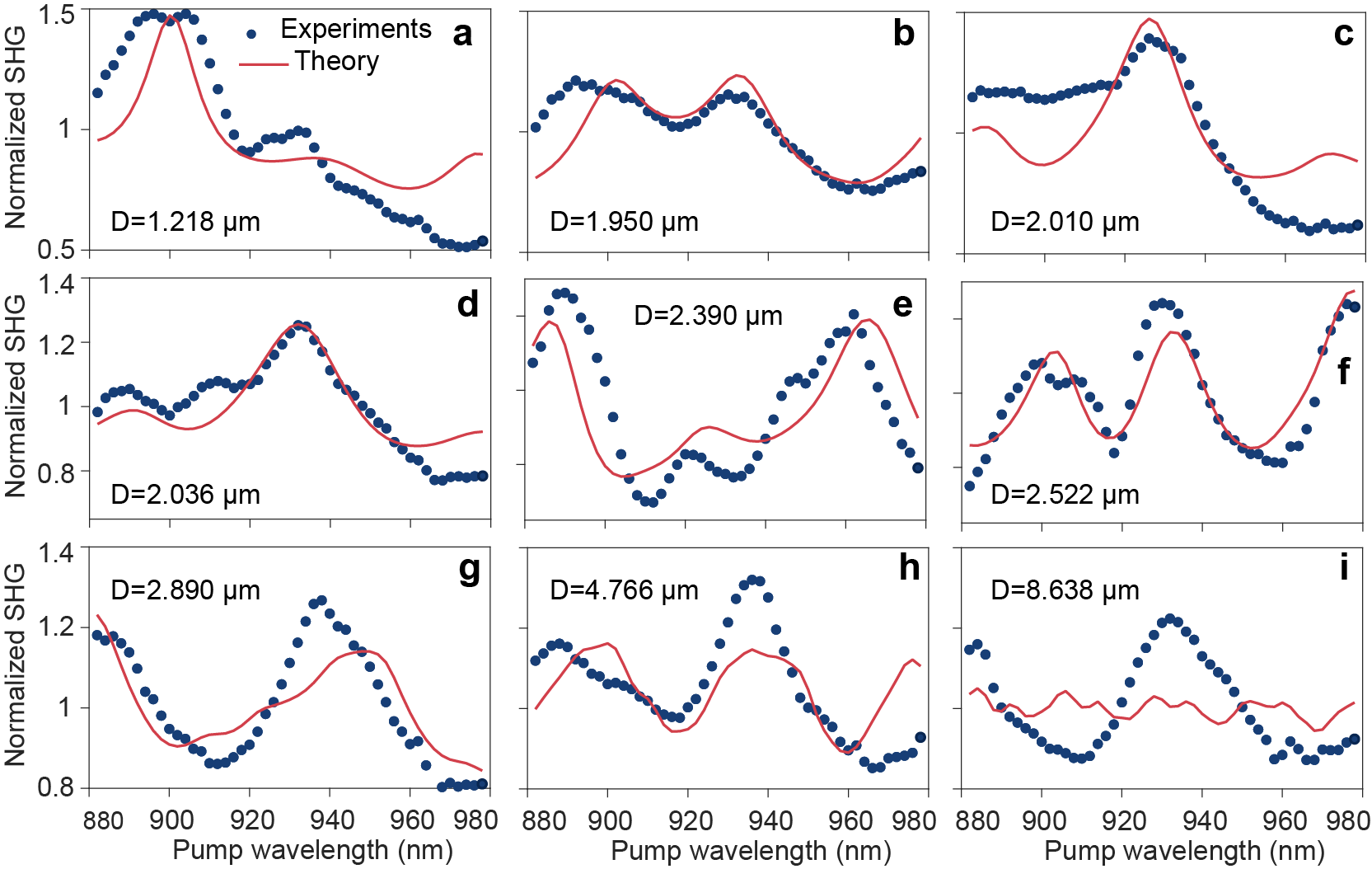}

   \caption{\textbf{Wavelength-dependent Mie-Driven SHG from the assembled micro-spheres}. \textbf{a-i}, SHG  measured from micro-spheres of different diameters obtained by sweeping the pump wavelength in-between 880-980~nm together with the best-fit function obtained by using Eq.~\ref{eq:equation1}, which considers the enhancement determined by the Mie resonances of the micro-spheres. Reported diameters have an error of $\pm$20~nm. Values of the the filling fractions and of the absorption coefficient returned by the fits are reported in the SI. Both experimental data and theory are normalized to their mean values.}
   \label{fig:figure5}
\end{figure*}
 We have measured the SHG (2$\omega$) on a set of 32 micro-spheres having diameters between 1-12~$\upmu$m, by illuminating them with pulses of 120~fs at 930~nm ($\omega$) in the configuration sketched in Fig.~\ref{fig:figure4}a (see Methods and supplementaries S6). The spectrum and the quadratic power-dependence of the nonlinear signal proves that we measured SHG (see supplementaries S7). SEM images of four micro-spheres with different diameters are shown in Fig.~\ref{fig:figure4}b-e and are considered as prototype examples to study size-dependent properties of the SHG. 
The images of the SHG reveal \textit{speckle} patterns due to the $\chi^{(2)}$ disorder, Fig.~\ref{fig:figure4}f-i, in striking contrast with the effective-medium behavior observed in the linear regime. 
Indeed, at 2$\omega$, there are multiple coherent emitters that interfere with random phase and amplitude.
The SHG comes from the entire micro-sphere, as shown by the size and the symmetry of the speckle images, which match those of the corresponding structures. 
The SHG appears more efficient in the center of the image, as a consequence of the spherical shape and consistently with the focusing of the pump. 
As shown in Fig.~\ref{fig:figure4}k-n,
the SHG is broadly and isotropically distributed over the solid angle of collection ($\approx$75$^\circ$), over an angular range much broader than that of the pump ($<$1$^\circ$). The efficiency is comparable along the different directions, with a more distinct emission at very large azimuthal angles ($>$60$^\circ$).
This effect can be addressed to the spherical shape and the size of the assembly, but further investigation is necessary to fully understand its origin.
The observed angular dispersion could be used to decouple the generated harmonic from the pump without the use of filters~\cite{fischer2006broadband}.  

The intensity distributions of the speckles are shown in Fig.~\ref{fig:figure4}j for three different micro-spheres. The observed Rayleigh distributions provide evidence of SHG interference from a large number of fully randomized emitters, without dominant orientation or position correlations~\cite{goodman1976some}. The exponential tail gets more pronounced for increasing diameters owing to the larger statistics available.
Also the SHG measured at variable input polarizations confirms the random nature of the $\chi^{(2)}$ disorder, Fig.~\ref{fig:figure4}o. Large micro-spheres (D$>$5$\upmu$m) show a fully isotropic response, corresponding to the  absence of polarization selection rules. When decreasing the size, the reduced number of domains leads to a partial averaging of the random interference and a residual polarization dependence appears. Also in this regime (D$<$5$\upmu$m), polarization selection rules are relaxed, with a variability of 20$\%$ on average.

Phase-matching properties are investigated in Fig.~\ref{fig:figure4}p. The SHG scales as the third power of the micro-sphere radius, reaching  diameters six times larger than the maximal coherence length of BaTiO$_3$ (see Supplementaries S1). The data correspond to a linear scaling over more than three orders of magnitude with the volume of the micro-spheres, i.e. with the number of the nano-crystals, unequivocally revealing RQPM.
The figure of merit for the conversion in the this regime is the comparison with a crystalline structure of the same size~\cite{baudrier2004random}. Due to the micron-size of our structures, we have performed the comparison numerically (see supplementaries S8). Results are shown in the inset of Fig.~\ref{fig:figure4}p.
Simulations reproduce the monotonic $\propto$L$^3$ scaling of RQPM for the disordered assemblies (blue line), pointing out that the SHG from the crystalline structures (gray lines) is destructively and constructively interfering in a periodic manner. Despite the extremely small scale, the SHG from the disordered assemblies is always comparable to that of the best-oriented crystalline counterparts (gray solid line), with the important advantage of having their phase-matching conditions relaxed. Interestingly, the disordered assembly largely outperforms a randomly oriented BaTiO$_3$ crystal (gray dashed line), showing an SHG that is two orders of magnitude stronger. This is a significant advantage, since controlling the lattice orientation, after or during the growth of the crystal, is a demanding task. Calculations of the absolute efficiency from the disordered assemblies is provided in the supplementaries S8.
%
\subsection*{Coupling to Mie resonances}
We investigated the wavelength-dependent SHG on several micro-spheres with increasing diameters,
by sweeping the pump laser over a 100~nm wavelength range around 930~nm. In Fig.~\ref{fig:figure5}a-i we observe a broadband tunability of the SHG, as expected by RQPM~\cite{fischer2006broadband}, additionally driven by the Mie resonances of the micro-spheres, which enhance the SHG at specific wavelengths. This resonant modulation in the RQPM regime is reported here for the first time.
To describe this remarkable observation we developed an extension of the RQPM model~\cite{vidal2006generation,chen2019non} to $\chi^{(2)}$disordered structures sustaining modes (see supplementaries S9). 
In such a case the specific pattern of the modes defines an inhomogeneous distribution of the pump  and of the SHG field. By assuming that the spatial features of the modes evolve on a scale larger than the mean size of the domain (large-scale modes), the fields can be  considered constant over the single domain. This assumption is well satisfied in the case of crystalline domains with a mean size of 50~nm, as in the experiments.
The SHG field from each crystalline domain is multiplied by the enhancement factor $\xi_i(\omega, 2\omega)=\omega^2 F_i^2(\omega)F_i(2\omega)$, where $F_i(\omega)$ and $F_i(2\omega)$ are the field enhancement of the pump and of the SHG respectively~\cite{zhang2014enhanced}.
Due to the $\chi^{(2)}$ disorder, the total SHG field is the result of a random walk in the complex plane of the SHG field, with a mode-dependent step-length distribution, as sketched in Fig.~S11.
The SHG intensity is given by the variance of this random walk and reads
\begin{equation}
   I_{2\omega} \propto \omega^2 E_\textrm{int}^2(\omega) E_\textrm{int}(2\omega) N
    \label{eq:equation1}
\end{equation}
 where $E_\textrm{int}$ is the total internal energy of the micro-sphere.
We used  Eq.\ref{eq:equation1} within a fitting procedure to describe the experiments, by calculating $E_\textrm{int}$  in the EEM approximation. Best-fit functions are reported in Fig.~\ref{fig:figure5} and are in good agreement with the measurements. The discrepancy observed for larger diameters is addressed to scattering inside the micro-spheres  and consequently to the limited validity of the EMM model. This is consistent with the discrepancies observed for the linear resonant scattering of the pump, although here we directly probe the internal energy of the micro-spheres and not the scattering cross-section. 
%
%
\section*{Conclusions}
We have realized a multifunctional platform based on the bottom-up assembly BaTiO$_3$ nano-crystals to study second-order nonlinear effects in disordered photonic materials. We observed an unexplored phase-matching mechanism that relies on the coupling of RQPM with the Mie resonances of the entire disordered structure. The assembled micro-spheres show a broadband and simultaneously Mie-enhanced SHG over a wavelength range of 100~nm. A main novelty is to have an optimization mechanism for the SHG at scales larger than the coherence length that is not related to the orientation and the quality of the crystal, but which exploits the geometry of the structure. Our model describes the main observations well  and pinpoints the essential physics from a complex interplay of linear and nonlinear effects. 

This work opens the door to a still uncharted research. Minimization of losses due to out-coupling with the substrate is of primary importance and, e.g. by introducing  a spacer, would drastically improve the Mie-resonant enhancement and potentially compensate for the weak efficiency of the nano-domains in comparison to the micro-domains. Other research development regards the use of a near-field coupling to excite specific modes and the minimization of the internal scattering in larger structures ($>$10$\upmu$m). A more elaborated model is necessary to explain the broad angular SHG emission and to estimate the contribution of overlapping modes in the RQPM regime. 
Since the SHG is a coherent process, there is a whole set of opportunities provided by disorder to achieve tailored interference~\cite{wiersma2013disordered,rotter2017light}, both for the pump and the SHG. Interesting perspectives are the hierarchical assembling of mono-dispersed or bi-dispersed micro-spheres into large-scale correlated structures~\cite{florescu2009designer} and the implementation of wave-front shaping  protocols to control and optimize the efficiency and the bandwidth of the nonlinear process. Thanks to the cost-effective and scalable production of the micro-spheres, we envision applications in the fabrication of tunable wide-acceptance-angle up-conversion screens~\cite{Barh2019upconversion}. Due to the time-reverse symmetry of the three-wave mixing process, results provided here could be extended to spontaneous parametric down-conversion (SPDC)~\cite{ru2017OPOrandom} and to the realization of disordered quantum sources.

\section*{Methods}
\subsection*{Emulsion-driven assembling}
Barium titanate (BaTiO$_3$) nano-crystals are purchased from Nyacol Nano Technology Inc.(BT80 25$\%$ wt.) as a stable aqueous dispersion.
The emulsion is prepared by vigorous hand-shaking of 10~$\upmu$l  dispersion (2$\%$ wt.) mixed with 2.5~ml of surfactant-loaded hexadecane (SPAN80 1$\%$ wt.). About 100~$\upmu$l of the emulsified mixture is transferred on a microscope glass slide before emulsion coalescence and is baked at 80~$^{\circ}$C for 12 hours in the oven. Glass slides are 1~mm thick. Deposited water-BaTiO$_3$ droplets are stable and tend to settle on a single layer over the slide. A complete water-into-hexadecane diffusion is already observed after 1 hour of baking. The remaining heat treatment is necessary to conclude the hexadecane evaporation. Left-over of the hexadecane and of the surfactant is removed by sequentially washing with hexane. Since BaTiO$_3$ has a a Curie temperature of 120~$^\circ$C the micro-spheres are not sintered. Exposing the nano-crystals to much higher temperatures could determine a non-reversible transition to the cubic crystalline phase and the loss of the non-centrosymmetry.
\subsection*{Effective-Medium-Mie model and fitting procedure}
The effective-medium-Mie (EEM) model first calculates the effective refractive index of the sphere through the Maxwell-Garnett mixing rule based on the average refractive index of bulk BaTiO$_3$  ($n\approx 2.4$ at 600~nm) and on a variable filling fraction. This value of $n_{\textrm{eff}}$ is used together with the diameters of the micro-sphere measured with the SEM  to calculate the unpolarized scattering cross section by using Mie theory for  spherical particles. 
We take the specific illumination-collection geometry of the measurements into account.
By varying the filling fraction of the micro-spheres between 45$\%$ and 65$\%$ and by fixing the absorption coefficient to $k=0.003$ we created a data-set for the fitting procedure. Both, experimental and theoretical spectra, are normalized to their mean values. Best-fitting curves are selected by using the least-square method. See Supplementaries S5 for more details. 

\subsection*{Effect of the substrate on Mie-resonances}
We simulated the scattering from a homogeneous sphere with effective refractive index $n_\mathrm{eff}=1.55$ placed on a glass substrate and illuminated by a plane wave at 930~nm, by using Comsol.
Results from the simulations are reported in the Supplementaries S5.3. These show that the substrate has little effect on the resonance positions, as it determines only a shift of a few nm. Differently, it introduces dominant out-coupling losses that smooth out resonant peaks. In the Supplementaries S5.3 we show that the presence of the substrate can be included in the analytical EMM theory by adding an imaginary part to the effective refractive index, with absorption coefficients typically of $k\approx 0.005$.

\subsection*{Optical nonlinear setup}
SHG from the micro-spheres has been investigated by free-space coupling to the samples with linearly polarized laser pulses of 120~fs at 930~nm ($\omega$), with a repetition rate of 80~MHz. We set the waist (FWHM) of the pump beam to 21~$\upmu$m, i.e. larger than the micro-spheres, to approximate a plane-wave illumination. The average power of the train of pulses was 180~mW, corresponding to a single-pulse fluence of 0.68~mJ/cm$^2$. The SHG (2$\omega$) was collected in the forward direction with a high-NA objective (NA=0.95, Apochromat). The angular emission of the SHG has been obtained by imaging the back-focal-plane (Fourier plane) of the collection objective. More details on the setup and the measurement procedures are in the Supplementaries S6.
\section*{Data availability}
The data that support the plots within this paper and other findings of this study are available from the corresponding authors upon reasonable request.
\section*{Code availability}
The codes that support the findings of this study are available from the corresponding authors upon reasonable request.
\newpage
\bibliography{main}

\begin{thebibliography}{10}
\urlstyle{rm}
\expandafter\ifx\csname url\endcsname\relax
  \def\url#1{\texttt{#1}}\fi
\expandafter\ifx\csname urlprefix\endcsname\relax\def\urlprefix{URL }\fi
\expandafter\ifx\csname doiprefix\endcsname\relax\def\doiprefix{DOI: }\fi
\providecommand{\bibinfo}[2]{#2}
\providecommand{\eprint}[2][]{\url{#2}}

\bibitem{boyd2008nonlinear}
\bibinfo{author}{Boyd, R.~W.}
\newblock \emph{\bibinfo{title}{Nonlinear Optics}}
  (\bibinfo{publisher}{Elsevier}, \bibinfo{year}{2008}).

\bibitem{KwiatZeilinger1995_PRL_PhotonPairs}
\bibinfo{author}{Kwiat, P.~G.} \emph{et~al.}
\newblock \bibinfo{journal}{\bibinfo{title}{New high-intensity source of
  polarization-entangled photon pairs}}.
\newblock {\emph{\JournalTitle{Phys. Rev. Lett.}}}
  \textbf{\bibinfo{volume}{75}}, \bibinfo{pages}{4337--4341}
  (\bibinfo{year}{1995}).

\bibitem{Shen:1996vl}
\bibinfo{author}{Shen, Y.~R.}
\newblock \bibinfo{journal}{\bibinfo{title}{{A few selected applications of
  surface nonlinear optical spectroscopy}}}.
\newblock {\emph{\JournalTitle{Proc Natl Acad Sci USA}}}
  \textbf{\bibinfo{volume}{93}}, \bibinfo{pages}{12104} (\bibinfo{year}{1996}).

\bibitem{Werner2015OL}
\bibinfo{author}{Werner, C.~S.}, \bibinfo{author}{Buse, K.} \&
  \bibinfo{author}{Breunig, I.}
\newblock \bibinfo{journal}{\bibinfo{title}{Continuous-wave whispering-gallery
  optical parametric oscillator for high-resolution spectroscopy}}.
\newblock {\emph{\JournalTitle{Opt. Lett.}}} \textbf{\bibinfo{volume}{40}},
  \bibinfo{pages}{772--775} (\bibinfo{year}{2015}).

\bibitem{campagnola2003second}
\bibinfo{author}{Campagnola, P.~J.} \& \bibinfo{author}{Loew, L.~M.}
\newblock \bibinfo{journal}{\bibinfo{title}{Second-harmonic imaging microscopy
  for visualizing biomolecular arrays in cells, tissues and organisms}}.
\newblock {\emph{\JournalTitle{Nature biotechnology}}}
  \textbf{\bibinfo{volume}{21}}, \bibinfo{pages}{1356} (\bibinfo{year}{2003}).

\bibitem{trebino2012frequency}
\bibinfo{author}{Trebino, R.}
\newblock \emph{\bibinfo{title}{Frequency-resolved optical gating: the
  measurement of ultrashort laser pulses}} (\bibinfo{publisher}{Springer
  Science \& Business Media}, \bibinfo{year}{2012}).

\bibitem{caspani2017integrated}
\bibinfo{author}{Caspani, L.} \emph{et~al.}
\newblock \bibinfo{journal}{\bibinfo{title}{Integrated sources of photon
  quantum states based on nonlinear optics}}.
\newblock {\emph{\JournalTitle{Light: Science \& Applications}}}
  \textbf{\bibinfo{volume}{6}}, \bibinfo{pages}{e17100} (\bibinfo{year}{2017}).

\bibitem{gunter2012nonlinear}
\bibinfo{author}{G{\"u}nter, P.}
\newblock \emph{\bibinfo{title}{Nonlinear optical effects and materials}},
  vol.~\bibinfo{volume}{72} (\bibinfo{publisher}{Springer},
  \bibinfo{year}{2012}).

\bibitem{fejer1992QPM}
\bibinfo{author}{Fejer, M.~M.}, \bibinfo{author}{Magel, G.},
  \bibinfo{author}{Jundt, D.~H.} \& \bibinfo{author}{Byer, R.~L.}
\newblock \bibinfo{journal}{\bibinfo{title}{Quasi-phase-matched second harmonic
  generation: tuning and tolerances}}.
\newblock {\emph{\JournalTitle{IEEE Journal of Quantum Electronics}}}
  \textbf{\bibinfo{volume}{28}}, \bibinfo{pages}{2631--2654}
  (\bibinfo{year}{1992}).

\bibitem{moutzouris2003modalPM}
\bibinfo{author}{Moutzouris, K.} \emph{et~al.}
\newblock \bibinfo{journal}{\bibinfo{title}{Second-harmonic generation through
  optimized modal phase matching in semiconductor waveguides}}.
\newblock {\emph{\JournalTitle{Applied physics letters}}}
  \textbf{\bibinfo{volume}{83}}, \bibinfo{pages}{620--622}
  (\bibinfo{year}{2003}).

\bibitem{lin2013cyclicPM}
\bibinfo{author}{Lin, G.}, \bibinfo{author}{F{\"u}rst, J.~U.},
  \bibinfo{author}{Strekalov, D.~V.} \& \bibinfo{author}{Yu, N.}
\newblock \bibinfo{journal}{\bibinfo{title}{Wide-range cyclic phase matching
  and second harmonic generation in whispering gallery resonators}}.
\newblock {\emph{\JournalTitle{Applied Physics Letters}}}
  \textbf{\bibinfo{volume}{103}}, \bibinfo{pages}{181107}
  (\bibinfo{year}{2013}).

\bibitem{rivoire2011cavity}
\bibinfo{author}{Rivoire, K.}, \bibinfo{author}{Buckley, S.} \&
  \bibinfo{author}{Vu{\v{c}}kovi{\'c}, J.}
\newblock \bibinfo{journal}{\bibinfo{title}{Multiply resonant photonic crystal
  nanocavities for nonlinear frequency conversion}}.
\newblock {\emph{\JournalTitle{Optics express}}} \textbf{\bibinfo{volume}{19}},
  \bibinfo{pages}{22198--22207} (\bibinfo{year}{2011}).

\bibitem{lin2016cavity}
\bibinfo{author}{Lin, Z.}, \bibinfo{author}{Liang, X.},
  \bibinfo{author}{Lon{\v{c}}ar, M.}, \bibinfo{author}{Johnson, S.~G.} \&
  \bibinfo{author}{Rodriguez, A.~W.}
\newblock \bibinfo{journal}{\bibinfo{title}{Cavity-enhanced second-harmonic
  generation via nonlinear-overlap optimization}}.
\newblock {\emph{\JournalTitle{Optica}}} \textbf{\bibinfo{volume}{3}},
  \bibinfo{pages}{233--238} (\bibinfo{year}{2016}).

\bibitem{suchowski2013phase}
\bibinfo{author}{Suchowski, H.} \emph{et~al.}
\newblock \bibinfo{journal}{\bibinfo{title}{Phase mismatch--free nonlinear
  propagation in optical zero-index materials}}.
\newblock {\emph{\JournalTitle{Science}}} \textbf{\bibinfo{volume}{342}},
  \bibinfo{pages}{1223--1226} (\bibinfo{year}{2013}).

\bibitem{pu2010coreshell}
\bibinfo{author}{Pu, Y.}, \bibinfo{author}{Grange, R.}, \bibinfo{author}{Hsieh,
  C.-L.} \& \bibinfo{author}{Psaltis, D.}
\newblock \bibinfo{journal}{\bibinfo{title}{Nonlinear optical properties of
  core-shell nanocavities for enhanced second-harmonic generation}}.
\newblock {\emph{\JournalTitle{Physical review letters}}}
  \textbf{\bibinfo{volume}{104}}, \bibinfo{pages}{207402}
  (\bibinfo{year}{2010}).

\bibitem{suchowski2010broadband}
\bibinfo{author}{Suchowski, H.}, \bibinfo{author}{Bruner, B.~D.},
  \bibinfo{author}{Arie, A.} \& \bibinfo{author}{Silberberg, Y.}
\newblock \bibinfo{journal}{\bibinfo{title}{Broadband nonlinear frequency
  conversion}}.
\newblock {\emph{\JournalTitle{Optics and Photonics News}}}
  \textbf{\bibinfo{volume}{21}}, \bibinfo{pages}{36--41}
  (\bibinfo{year}{2010}).

\bibitem{baudrier2004random}
\bibinfo{author}{Baudrier-Raybaut, M.}, \bibinfo{author}{Haidar, R.},
  \bibinfo{author}{Kupecek, P.}, \bibinfo{author}{Lemasson, P.} \&
  \bibinfo{author}{Rosencher, E.}
\newblock \bibinfo{journal}{\bibinfo{title}{Random quasi-phase-matching in bulk
  polycrystalline isotropic nonlinear materials}}.
\newblock {\emph{\JournalTitle{Nature}}} \textbf{\bibinfo{volume}{432}},
  \bibinfo{pages}{374} (\bibinfo{year}{2004}).

\bibitem{skipetrov2004RQPM}
\bibinfo{author}{Skipetrov, S.~E.}
\newblock \bibinfo{journal}{\bibinfo{title}{Nonlinear optics: Disorder is the
  new order}}.
\newblock {\emph{\JournalTitle{Nature}}} \textbf{\bibinfo{volume}{432}},
  \bibinfo{pages}{285} (\bibinfo{year}{2004}).

\bibitem{vidal2006generation}
\bibinfo{author}{Vidal, X.} \& \bibinfo{author}{Martorell, J.}
\newblock \bibinfo{journal}{\bibinfo{title}{Generation of light in media with a
  random distribution of nonlinear domains}}.
\newblock {\emph{\JournalTitle{Physical review letters}}}
  \textbf{\bibinfo{volume}{97}}, \bibinfo{pages}{013902}
  (\bibinfo{year}{2006}).

\bibitem{bravo2010optical}
\bibinfo{author}{Bravo-Abad, J.}, \bibinfo{author}{Vidal, X.},
  \bibinfo{author}{Ju{\'a}rez, J. L.~D.} \& \bibinfo{author}{Martorell, J.}
\newblock \bibinfo{journal}{\bibinfo{title}{Optical second-harmonic scattering
  from a non-diffusive random distribution of nonlinear domains}}.
\newblock {\emph{\JournalTitle{Optics express}}} \textbf{\bibinfo{volume}{18}},
  \bibinfo{pages}{14202--14211} (\bibinfo{year}{2010}).

\bibitem{ru2017OPOrandom}
\bibinfo{author}{Ru, Q.} \emph{et~al.}
\newblock \bibinfo{journal}{\bibinfo{title}{Optical parametric oscillation in a
  random polycrystalline medium}}.
\newblock {\emph{\JournalTitle{Optica}}} \textbf{\bibinfo{volume}{4}},
  \bibinfo{pages}{617--618} (\bibinfo{year}{2017}).

\bibitem{fischer2006broadband}
\bibinfo{author}{Fischer, R.}, \bibinfo{author}{Saltiel, S.},
  \bibinfo{author}{Neshev, D.}, \bibinfo{author}{Krolikowski, W.} \&
  \bibinfo{author}{Kivshar, Y.~S.}
\newblock \bibinfo{journal}{\bibinfo{title}{Broadband femtosecond frequency
  doubling in random media}}.
\newblock {\emph{\JournalTitle{Applied physics letters}}}
  \textbf{\bibinfo{volume}{89}}, \bibinfo{pages}{191105}
  (\bibinfo{year}{2006}).

\bibitem{molina2008NLphotglass}
\bibinfo{author}{Molina, P.}, \bibinfo{author}{Ramírez, M. d. l.~O.} \&
  \bibinfo{author}{Bausá, L.~E.}
\newblock \bibinfo{journal}{\bibinfo{title}{Strontium barium niobate as a
  multifunctional two-dimensional nonlinear “photonic glass”}}.
\newblock {\emph{\JournalTitle{Advanced Functional Materials}}}
  \textbf{\bibinfo{volume}{18}}, \bibinfo{pages}{709--715}
  (\bibinfo{year}{2008}).

\bibitem{ricciardi2015frequency}
\bibinfo{author}{Ricciardi, I.} \emph{et~al.}
\newblock \bibinfo{journal}{\bibinfo{title}{Frequency comb generation in
  quadratic nonlinear media}}.
\newblock {\emph{\JournalTitle{Physical Review A}}}
  \textbf{\bibinfo{volume}{91}}, \bibinfo{pages}{063839}
  (\bibinfo{year}{2015}).

\bibitem{kuznetsov2016optically}
\bibinfo{author}{Kuznetsov, A.~I.}, \bibinfo{author}{Miroshnichenko, A.~E.},
  \bibinfo{author}{Brongersma, M.~L.}, \bibinfo{author}{Kivshar, Y.~S.} \&
  \bibinfo{author}{Luk’yanchuk, B.}
\newblock \bibinfo{journal}{\bibinfo{title}{Optically resonant dielectric
  nanostructures}}.
\newblock {\emph{\JournalTitle{Science}}} \textbf{\bibinfo{volume}{354}},
  \bibinfo{pages}{aag2472} (\bibinfo{year}{2016}).

\bibitem{kim2013second}
\bibinfo{author}{Kim, E.} \emph{et~al.}
\newblock \bibinfo{journal}{\bibinfo{title}{Second-harmonic generation of
  single batio3 nanoparticles down to 22 nm diameter}}.
\newblock {\emph{\JournalTitle{ACS nano}}} \textbf{\bibinfo{volume}{7}},
  \bibinfo{pages}{5343--5349} (\bibinfo{year}{2013}).

\bibitem{timpu2016second}
\bibinfo{author}{Timpu, F.}, \bibinfo{author}{Sergeyev, A.},
  \bibinfo{author}{Hendricks, N.~R.} \& \bibinfo{author}{Grange, R.}
\newblock \bibinfo{journal}{\bibinfo{title}{Second-harmonic enhancement with
  mie resonances in perovskite nanoparticles}}.
\newblock {\emph{\JournalTitle{Acs Photonics}}} \textbf{\bibinfo{volume}{4}},
  \bibinfo{pages}{76--84} (\bibinfo{year}{2016}).

\bibitem{deBoer1993SHGcorr}
\bibinfo{author}{de~Boer, J.~F.}, \bibinfo{author}{Lagendijk, A.},
  \bibinfo{author}{Sprik, R.} \& \bibinfo{author}{Feng, S.}
\newblock \bibinfo{journal}{\bibinfo{title}{Transmission and reflection
  correlations of second harmonic waves in nonlinear random media}}.
\newblock {\emph{\JournalTitle{Physical review letters}}}
  \textbf{\bibinfo{volume}{71}}, \bibinfo{pages}{3947} (\bibinfo{year}{1993}).

\bibitem{faez2009SHGdiff}
\bibinfo{author}{Faez, S.}, \bibinfo{author}{Johnson, P.},
  \bibinfo{author}{Mazurenko, D.} \& \bibinfo{author}{Lagendijk, A.}
\newblock \bibinfo{journal}{\bibinfo{title}{Experimental observation of
  second-harmonic generation and diffusion inside random media}}.
\newblock {\emph{\JournalTitle{JOSA B}}} \textbf{\bibinfo{volume}{26}},
  \bibinfo{pages}{235--243} (\bibinfo{year}{2009}).

\bibitem{makeev2003second}
\bibinfo{author}{Makeev, E.} \& \bibinfo{author}{Skipetrov, S.}
\newblock \bibinfo{journal}{\bibinfo{title}{Second harmonic generation in
  suspensions of spherical particles}}.
\newblock {\emph{\JournalTitle{Optics communications}}}
  \textbf{\bibinfo{volume}{224}}, \bibinfo{pages}{139--147}
  (\bibinfo{year}{2003}).

\bibitem{kim2008microspheres}
\bibinfo{author}{Kim, S.-H.} \emph{et~al.}
\newblock \bibinfo{journal}{\bibinfo{title}{Microspheres with tunable
  refractive index by controlled assembly of nanoparticles}}.
\newblock {\emph{\JournalTitle{Advanced Materials}}}
  \textbf{\bibinfo{volume}{20}}, \bibinfo{pages}{3268--3273}
  (\bibinfo{year}{2008}).

\bibitem{vogel2015color}
\bibinfo{author}{Vogel, N.} \emph{et~al.}
\newblock \bibinfo{journal}{\bibinfo{title}{Color from hierarchy: Diverse
  optical properties of micron-sized spherical colloidal assemblies}}.
\newblock {\emph{\JournalTitle{Proceedings of the National Academy of
  Sciences}}} \textbf{\bibinfo{volume}{112}}, \bibinfo{pages}{10845--10850}
  (\bibinfo{year}{2015}).

\bibitem{VoglerNeuling_pssb2020}
\bibinfo{author}{Vogler-Neuling, V.~V.} \emph{et~al.}
\newblock \bibinfo{journal}{\bibinfo{title}{Solution-processed barium titanate
  nonlinear woodpile photonic structures with large surface areas}}.
\newblock {\emph{\JournalTitle{physica status solidi (b)}}}
  \bibinfo{pages}{1900755} (\bibinfo{year}{2020}).

\bibitem{yang2014super}
\bibinfo{author}{Yang, H.}, \bibinfo{author}{Moullan, N.},
  \bibinfo{author}{Auwerx, J.} \& \bibinfo{author}{Gijs, M.~A.}
\newblock \bibinfo{journal}{\bibinfo{title}{Super-resolution biological
  microscopy using virtual imaging by a microsphere nanoscope}}.
\newblock {\emph{\JournalTitle{Small}}} \textbf{\bibinfo{volume}{10}},
  \bibinfo{pages}{1712--1718} (\bibinfo{year}{2014}).

\bibitem{checcucci2018titania}
\bibinfo{author}{Checcucci, S.} \emph{et~al.}
\newblock \bibinfo{journal}{\bibinfo{title}{Titania-based spherical mie
  resonators elaborated by high-throughput aerosol spray: Single object
  investigation}}.
\newblock {\emph{\JournalTitle{Advanced Functional Materials}}}
  \textbf{\bibinfo{volume}{28}}, \bibinfo{pages}{1801958}
  (\bibinfo{year}{2018}).

\bibitem{lalanne2018light}
\bibinfo{author}{Lalanne, P.}, \bibinfo{author}{Yan, W.},
  \bibinfo{author}{Vynck, K.}, \bibinfo{author}{Sauvan, C.} \&
  \bibinfo{author}{Hugonin, J.-P.}
\newblock \bibinfo{journal}{\bibinfo{title}{Light interaction with photonic and
  plasmonic resonances}}.
\newblock {\emph{\JournalTitle{Laser \& Photonics Reviews}}}
  \textbf{\bibinfo{volume}{12}}, \bibinfo{pages}{1700113}
  (\bibinfo{year}{2018}).

\bibitem{chen2004photonic}
\bibinfo{author}{Chen, Z.}, \bibinfo{author}{Taflove, A.} \&
  \bibinfo{author}{Backman, V.}
\newblock \bibinfo{journal}{\bibinfo{title}{Photonic nanojet enhancement of
  backscattering of light by nanoparticles: a potential novel visible-light
  ultramicroscopy technique}}.
\newblock {\emph{\JournalTitle{Optics express}}} \textbf{\bibinfo{volume}{12}},
  \bibinfo{pages}{1214--1220} (\bibinfo{year}{2004}).

\bibitem{Geints2012_nanojet}
\bibinfo{author}{Geints, Y.~E.}, \bibinfo{author}{Zemlyanov, A.~A.} \&
  \bibinfo{author}{Panina, E.~K.}
\newblock \bibinfo{journal}{\bibinfo{title}{Photonic jets from resonantly
  excited transparent dielectric microspheres}}.
\newblock {\emph{\JournalTitle{J. Opt. Soc. Am. B}}}
  \textbf{\bibinfo{volume}{29}}, \bibinfo{pages}{758--762}
  (\bibinfo{year}{2012}).

\bibitem{goodman1976some}
\bibinfo{author}{Goodman, J.~W.}
\newblock \bibinfo{journal}{\bibinfo{title}{Some fundamental properties of
  speckle}}.
\newblock {\emph{\JournalTitle{JOSA}}} \textbf{\bibinfo{volume}{66}},
  \bibinfo{pages}{1145--1150} (\bibinfo{year}{1976}).

\bibitem{chen2019non}
\bibinfo{author}{Chen, X.} \& \bibinfo{author}{Gaume, R.}
\newblock \bibinfo{journal}{\bibinfo{title}{Non-stoichiometric grain-growth in
  znse ceramics for $\chi$ (2) interaction}}.
\newblock {\emph{\JournalTitle{Optical Materials Express}}}
  \textbf{\bibinfo{volume}{9}}, \bibinfo{pages}{400--409}
  (\bibinfo{year}{2019}).

\bibitem{zhang2014enhanced}
\bibinfo{author}{Zhang, J.}, \bibinfo{author}{Cassan, E.} \&
  \bibinfo{author}{Zhang, X.}
\newblock \bibinfo{journal}{\bibinfo{title}{Enhanced mid-to-near-infrared
  second harmonic generation in silicon plasmonic microring resonators with low
  pump power}}.
\newblock {\emph{\JournalTitle{Photonics Research}}}
  \textbf{\bibinfo{volume}{2}}, \bibinfo{pages}{143--149}
  (\bibinfo{year}{2014}).

\bibitem{wiersma2013disordered}
\bibinfo{author}{Wiersma, D.~S.}
\newblock \bibinfo{journal}{\bibinfo{title}{Disordered photonics}}.
\newblock {\emph{\JournalTitle{Nature Photonics}}}
  \textbf{\bibinfo{volume}{7}}, \bibinfo{pages}{188} (\bibinfo{year}{2013}).

\bibitem{rotter2017light}
\bibinfo{author}{Rotter, S.} \& \bibinfo{author}{Gigan, S.}
\newblock \bibinfo{journal}{\bibinfo{title}{Light fields in complex media:
  Mesoscopic scattering meets wave control}}.
\newblock {\emph{\JournalTitle{Reviews of Modern Physics}}}
  \textbf{\bibinfo{volume}{89}}, \bibinfo{pages}{015005}
  (\bibinfo{year}{2017}).

\bibitem{florescu2009designer}
\bibinfo{author}{Florescu, M.}, \bibinfo{author}{Torquato, S.} \&
  \bibinfo{author}{Steinhardt, P.~J.}
\newblock \bibinfo{journal}{\bibinfo{title}{Designer disordered materials with
  large, complete photonic band gaps}}.
\newblock {\emph{\JournalTitle{Proceedings of the National Academy of
  Sciences}}} \textbf{\bibinfo{volume}{106}}, \bibinfo{pages}{20658--20663}
  (\bibinfo{year}{2009}).

\bibitem{Barh2019upconversion}
\bibinfo{author}{Barh, A.}, \bibinfo{author}{Rodrigo, P.~J.},
  \bibinfo{author}{Meng, L.}, \bibinfo{author}{Pedersen, C.} \&
  \bibinfo{author}{Tidemand-Lichtenberg, P.}
\newblock \bibinfo{journal}{\bibinfo{title}{Parametric upconversion imaging and
  its applications}}.
\newblock {\emph{\JournalTitle{Adv. Opt. Photon.}}}
  \textbf{\bibinfo{volume}{11}}, \bibinfo{pages}{952--1019}
  (\bibinfo{year}{2019}).

\end{thebibliography}


\begin{thebibliography}{10}
\urlstyle{rm}
\expandafter\ifx\csname url\endcsname\relax
  \def\url#1{\texttt{#1}}\fi
\expandafter\ifx\csname urlprefix\endcsname\relax\def\urlprefix{URL }\fi
\expandafter\ifx\csname doiprefix\endcsname\relax\def\doiprefix{DOI: }\fi
\providecommand{\bibinfo}[2]{#2}
\providecommand{\eprint}[2][]{\url{#2}}

\bibitem{boyd2008nonlinear}
\bibinfo{author}{Boyd, R.~W.}
\newblock \emph{\bibinfo{title}{Nonlinear Optics}}
  (\bibinfo{publisher}{Elsevier}, \bibinfo{year}{2008}).

\bibitem{zelmon1997refractive}
\bibinfo{author}{Zelmon, D.~E.}, \bibinfo{author}{Small, D.~L.} \&
  \bibinfo{author}{Schunemann, P.}
\newblock \bibinfo{journal}{\bibinfo{title}{Refractive index measurements of
  barium titanate from. 4 to 5.0 microns and implications for periodically
  poled frequency conversion devices}}.
\newblock {\emph{\JournalTitle{MRS Online Proceedings Library Archive}}}
  \textbf{\bibinfo{volume}{484}} (\bibinfo{year}{1997}).

\bibitem{zhao2000stress}
\bibinfo{author}{Zhao, T.}, \bibinfo{author}{Lu, H.}, \bibinfo{author}{Chen,
  F.}, \bibinfo{author}{Yang, G.} \& \bibinfo{author}{Chen, Z.}
\newblock \bibinfo{journal}{\bibinfo{title}{Stress-induced enhancement of
  second-order nonlinear optical susceptibilities of barium titanate films}}.
\newblock {\emph{\JournalTitle{Journal of Applied Physics}}}
  \textbf{\bibinfo{volume}{87}}, \bibinfo{pages}{7448--7451}
  (\bibinfo{year}{2000}).

\bibitem{otsu1979threshold}
\bibinfo{author}{Otsu, N.}
\newblock \bibinfo{journal}{\bibinfo{title}{A threshold selection method from
  gray-level histograms}}.
\newblock {\emph{\JournalTitle{IEEE transactions on systems, man, and
  cybernetics}}} \textbf{\bibinfo{volume}{9}}, \bibinfo{pages}{62--66}
  (\bibinfo{year}{1979}).

\bibitem{zurcher2018evaporation}
\bibinfo{author}{Z{\"u}rcher, J.}, \bibinfo{author}{Burg, B.~R.},
  \bibinfo{author}{Del~Carro, L.}, \bibinfo{author}{Studart, A.~R.} \&
  \bibinfo{author}{Brunschwiler, T.}
\newblock \bibinfo{journal}{\bibinfo{title}{On the evaporation of colloidal
  suspensions in confined pillar arrays}}.
\newblock {\emph{\JournalTitle{Transport in Porous Media}}}
  \textbf{\bibinfo{volume}{125}}, \bibinfo{pages}{173--192}
  (\bibinfo{year}{2018}).

\bibitem{zhuromskyy2017applicability}
\bibinfo{author}{Zhuromskyy, O.}
\newblock \bibinfo{journal}{\bibinfo{title}{Applicability of effective medium
  approximations to modelling of mesocrystal optical properties}}.
\newblock {\emph{\JournalTitle{Crystals}}} \textbf{\bibinfo{volume}{7}},
  \bibinfo{pages}{1} (\bibinfo{year}{2017}).

\bibitem{voshchinnikov2007effective}
\bibinfo{author}{Voshchinnikov, N.~V.}, \bibinfo{author}{Videen, G.} \&
  \bibinfo{author}{Henning, T.}
\newblock \bibinfo{journal}{\bibinfo{title}{Effective medium theories for
  irregular fluffy structures: aggregation of small particles}}.
\newblock {\emph{\JournalTitle{Applied Optics}}} \textbf{\bibinfo{volume}{46}},
  \bibinfo{pages}{4065--4072} (\bibinfo{year}{2007}).

\bibitem{markel2016introduction}
\bibinfo{author}{Markel, V.~A.}
\newblock \bibinfo{journal}{\bibinfo{title}{Introduction to the maxwell garnett
  approximation: tutorial}}.
\newblock {\emph{\JournalTitle{JOSA A}}} \textbf{\bibinfo{volume}{33}},
  \bibinfo{pages}{1244--1256} (\bibinfo{year}{2016}).

\bibitem{vettenburg2019calculating}
\bibinfo{author}{Vettenburg, T.}, \bibinfo{author}{Horsley, S.~A.} \&
  \bibinfo{author}{Bertolotti, J.}
\newblock \bibinfo{journal}{\bibinfo{title}{Calculating coherent light-wave
  propagation in large heterogeneous media}}.
\newblock {\emph{\JournalTitle{Optics express}}} \textbf{\bibinfo{volume}{27}},
  \bibinfo{pages}{11946--11967} (\bibinfo{year}{2019}).

\bibitem{bohren2008absorption}
\bibinfo{author}{Bohren, C.~F.} \& \bibinfo{author}{Huffman, D.~R.}
\newblock \emph{\bibinfo{title}{Absorption and scattering of light by small
  particles}} (\bibinfo{publisher}{John Wiley \& Sons}, \bibinfo{year}{2008}).

\bibitem{matzler2002matlab}
\bibinfo{author}{M{\"a}tzler, C.}
\newblock \bibinfo{journal}{\bibinfo{title}{Matlab functions for mie scattering
  and absorption, version 2}}.
\newblock {\emph{\JournalTitle{IAP Res. Rep}}} \textbf{\bibinfo{volume}{8}},
  \bibinfo{pages}{9} (\bibinfo{year}{2002}).

\bibitem{chen2019non}
\bibinfo{author}{Chen, X.} \& \bibinfo{author}{Gaume, R.}
\newblock \bibinfo{journal}{\bibinfo{title}{Non-stoichiometric grain-growth in
  znse ceramics for $\chi$ (2) interaction}}.
\newblock {\emph{\JournalTitle{Optical Materials Express}}}
  \textbf{\bibinfo{volume}{9}}, \bibinfo{pages}{400--409}
  (\bibinfo{year}{2019}).

\bibitem{vidal2006generation}
\bibinfo{author}{Vidal, X.} \& \bibinfo{author}{Martorell, J.}
\newblock \bibinfo{journal}{\bibinfo{title}{Generation of light in media with a
  random distribution of nonlinear domains}}.
\newblock {\emph{\JournalTitle{Physical review letters}}}
  \textbf{\bibinfo{volume}{97}}, \bibinfo{pages}{013902}
  (\bibinfo{year}{2006}).

\bibitem{baudrier2004random}
\bibinfo{author}{Baudrier-Raybaut, M.}, \bibinfo{author}{Haidar, R.},
  \bibinfo{author}{Kupecek, P.}, \bibinfo{author}{Lemasson, P.} \&
  \bibinfo{author}{Rosencher, E.}
\newblock \bibinfo{journal}{\bibinfo{title}{Random quasi-phase-matching in bulk
  polycrystalline isotropic nonlinear materials}}.
\newblock {\emph{\JournalTitle{Nature}}} \textbf{\bibinfo{volume}{432}},
  \bibinfo{pages}{374} (\bibinfo{year}{2004}).

\bibitem{zhang2014enhanced}
\bibinfo{author}{Zhang, J.}, \bibinfo{author}{Cassan, E.} \&
  \bibinfo{author}{Zhang, X.}
\newblock \bibinfo{journal}{\bibinfo{title}{Enhanced mid-to-near-infrared
  second harmonic generation in silicon plasmonic microring resonators with low
  pump power}}.
\newblock {\emph{\JournalTitle{Photonics Research}}}
  \textbf{\bibinfo{volume}{2}}, \bibinfo{pages}{143--149}
  (\bibinfo{year}{2014}).

\bibitem{lin2016cavity}
\bibinfo{author}{Lin, Z.}, \bibinfo{author}{Liang, X.},
  \bibinfo{author}{Lon{\v{c}}ar, M.}, \bibinfo{author}{Johnson, S.~G.} \&
  \bibinfo{author}{Rodriguez, A.~W.}
\newblock \bibinfo{journal}{\bibinfo{title}{Cavity-enhanced second-harmonic
  generation via nonlinear-overlap optimization}}.
\newblock {\emph{\JournalTitle{Optica}}} \textbf{\bibinfo{volume}{3}},
  \bibinfo{pages}{233--238} (\bibinfo{year}{2016}).

\end{thebibliography}

\section*{Acknowledgements}
The authors acknowledge fruitful discussions with Marc Lehner, Claude Renaut and Viola Vogler-Neuling. 
The authors acknowledge support from the FIRST—Center for Micro and Nanoscience of ETHZ and from the Scientific Center of Optical and Electron Microscopy (ScopeM) of ETHZ.
The project has received funding from the European
Union’s Horizon 2020 research and innovation program under the
Marie Skłodowska-Curie grant agreement No. 800487 (SECOONDO) and from the
European Research Council under the Grant Agreement No. 714837
(Chi2-nano-oxides).
The authors thank the Swiss National Science Foundation
(SNF) grants 150609. 
\section*{Authors contributions}
R.S. and R.G. conceived the work. R.S. M.Z. and L.I. developed the assembly method. R.S, A.M., F.K. and M.R.E. realized the structures, performed the FIB cuts and the SEM characterization. A.M., J.S.M. and F.T. performed the simulations. R.S. and A.M. developed the theory. R.S., A.M., J.S.M. and R.G. analyzed the data. R.S. wrote the first draft of the manuscript. All authors discussed the results and contributed to the writing of the manuscript.   
\section*{Competing interests}
The authors declare no competing interests.

\end{document}


\begin{singlespace} 

\maketitle

\noindent 

\tableofcontents
\thispagestyle{empty}
\end{singlespace}
\newpage
%
%

\section{The material: Barium Titanate (BaTiO$_3$)}
\label{sec:thematerial}
BaTiO$_3$ is particularly attractive for photonic applications since it has a high refractive index, a high second-order nonlinearity, a wide transparency window and several functional properties as electro-optic, elasto-optic and thermo-optic effects. Here, we describe linear and nonlinear optical properties relevant for this work. 
\subsection{Refractive index and $\chi^{(2)}$ tensor }
\begin{figure}[h]
    \centering
    \includegraphics[scale=1]{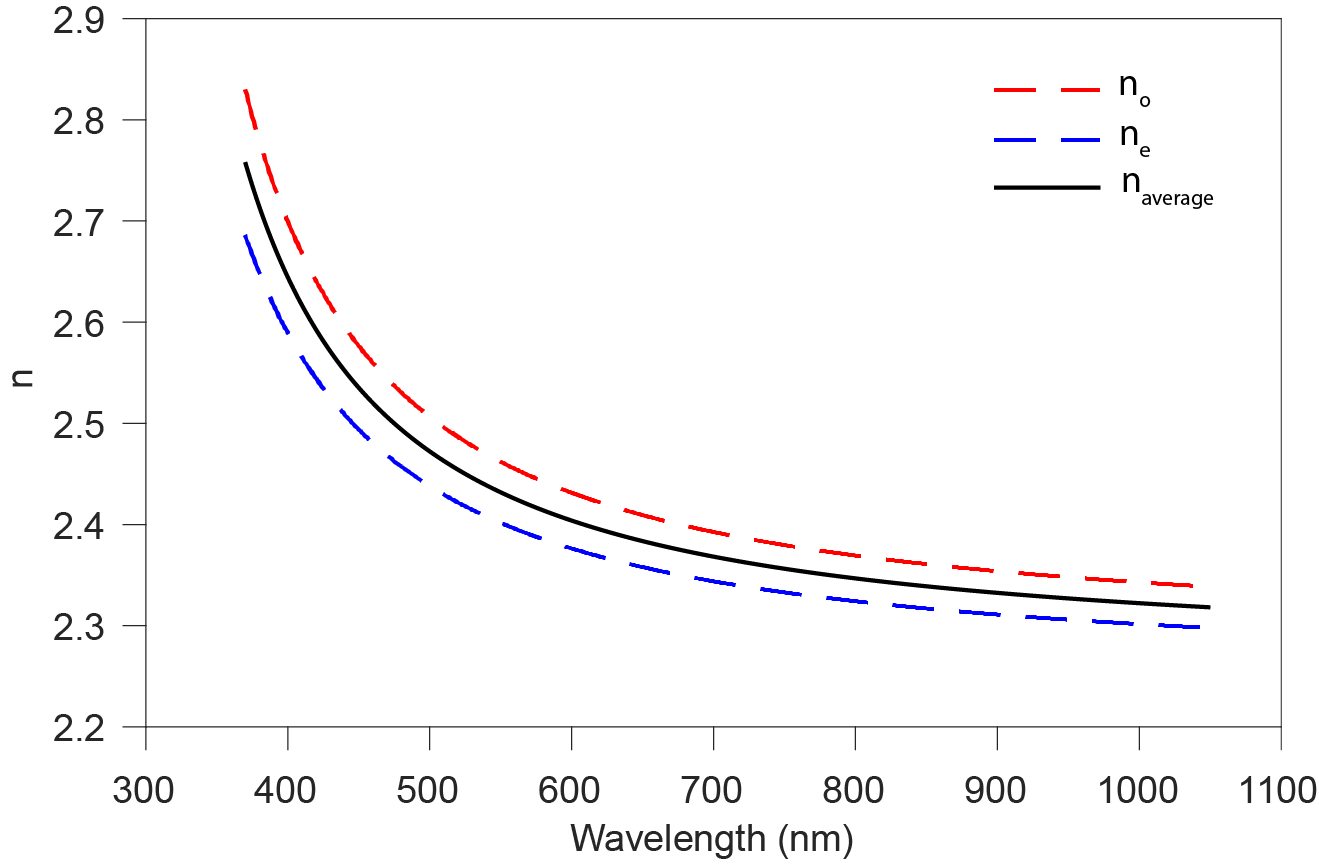} 
    \caption{Spectral dependence of ordinary, extraordinary and average refractive indices for bulk BaTiO$_3$}
	\label{fig:n_BTO}%
\end{figure}

BaTiO$_3$ is a (negative) uniaxial  birefringent crystal, hence its extraordinary index $n_\textrm{e}$ depends on the angle between $\vec{k}$ and the optic axis according to
Eq.~\ref{eq:refractive-index-angledependent}~\cite{boyd2008nonlinear}
\begin{equation}
\label{eq:refractive-index-angledependent}
\frac{1}{n(\theta)^2} = \frac{\sin^2\theta}{n_\textrm{e}^2} + \frac{\cos^2\theta}{n_o^2},
\end{equation}
with a maximal value $n_\textrm{e}(\theta = \pi) = n_\textrm{e}$ for $\theta = \pi$ and a minimal value $n_\textrm{e}(\theta = 0) = n_o$.  
The ordinary index $n_o$ , the maximal extraordinary index $n_\textrm{e}$ and their average value $n_\textrm{average}=(n_\textrm{o}+n_\textrm{e})/2$ are calculated at different wavelengths according to Zelmon D. et al.~\cite{zelmon1997refractive} and are shown in Fig.~\ref{fig:n_BTO}. For example, at 600~nm  one has $n_\textrm{average}=2.40$, while at 930~nm  one has $n_\textrm{average}=2.33$.  
 
The components of the  $\chi^{(2)}$ susceptibility tensor are reported in Eq.~\ref{eq:BTO_tensor} according to Tong Zhao et al.~\cite{zhao2000stress}
\begin{equation}
\label{eq:BTO_tensor}
d_\mathrm{BTO} =
\begin{bmatrix} 
0 & 0 & 0 & 0 & d_{15} & 0 \\
0 & 0 & 0 & d_{15} & 0 & 0 \\
d_{31} & d_{31} & d_{33} & 0 & 0 & 0 \\ 
\end{bmatrix}
\quad\textrm{with}\quad
\begin{matrix*}[r]
d_{15} = &-17.0 & \text{ pm/V}\\
d_{31} = &-15.7 & \text{ pm/V}\\
d_{33} = &- 6.8 & \text{ pm/V}\\
\end{matrix*}
\end{equation}

\subsection{Calculation of the coherence length}
\label{}
Due to the dispersion of $n$, two beams at $\omega_1$ and $\omega_2$ do not travel through the medium at the same speed, creating the phase mismatch $\Delta k$:
\begin{equation}
	\label{eq:phase-mismatch}
	\Delta k = 2k_1 - k_2 = \frac{2\omega_1}{c}(n(\omega_1)-n(\omega_2))
\end{equation}
For the case of SHG, this leads to a phase shift between two SHG waves generated at different points within the crystal. The length at which the relative phase shift of the generated waves reaches $\frac{\pi}{2}$ is called the coherence length $L_c$:

\begin{equation}
	\label{eq:coherence-length}
	L_c = \frac{\pi}{\Delta k}
\end{equation}
After one coherence length within the crystal, the SHG starts decreasing until it reaches zero at two coherence lengths, where there is total destructive interference between all SHG contributions. For this reason, the maximum possible SHG in a single (non-phase-matched) crystal is reached at lengths equal to an odd multiple of the coherence length, and minima at even multiples. 

In order to calculate the phase mismatch $\Delta k$ between the fundamental frequency and the second harmonic it is required to know the four different k-vectors (ordinary and extraordinary part of the fundamental and the SHG)  $k_o(\omega), k_e(\omega), k_o(2\omega), k_e(2\omega)$, which are found from $k = \frac{\omega n}{c}$ using the respective refractive index. The phase mismatch (defined in Eq.~\ref{eq:phase-mismatch}) is more generally  given as  $\Delta k = k_1 + k_2 - k_3$, where $k_1$ and $k_2$ are the fundamental frequencies and $k_3$ is the k-vector of the second harmonic (Boyd\cite{boyd2008nonlinear}, p.76). This leads us to eight different values for $\Delta k$ depending on the beam combinations:
\begin{equation}
	\begin{aligned}
		\Delta k^{ooo} = k_o(\omega) + k_o(\omega) - k_o(2\omega) \\
		\Delta k^{eeo} = k_e(\omega) + k_e(\omega) - k_o(2\omega) \\
		\Delta k^{oeo} = k_o(\omega) + k_e(\omega) - k_o(2\omega) \\
		\Delta k^{eoo} = k_o(\omega) + k_e(\omega) - k_o(2\omega) \\
		\Delta k^{ooe} = k_o(\omega) + k_o(\omega) - k_e(2\omega) \\
		\Delta k^{eee} = k_e(\omega) + k_e(\omega) - k_e(2\omega) \\
		\Delta k^{oee} = k_o(\omega) + k_e(\omega) - k_e(2\omega) \\
		\Delta k^{eoe} = k_e(\omega) + k_o(\omega) - k_e(2\omega) \\
	\end{aligned}
\end{equation}
 From this, the coherence length can be calculated according to  Eq.~\ref{eq:coherence-length}.  While beam combinations of type $ooe$  will see the largest coherence length, the beam combination $ooo$ has an angle independent coherence length. The calculated coherence length of BaTiO$_3$ at 930~nm  lies within the interval [0.95~$\upmu$m,1.87~$\upmu$m], with a constant $ooo$-value of 1.57~$\upmu$m and an average of 1.23~$\upmu$m.
%
\section{Assembled BaTiO$_3$ micro-spheres of different sizes}
\begin{figure}[h]
    \centering
    \includegraphics[scale=1]{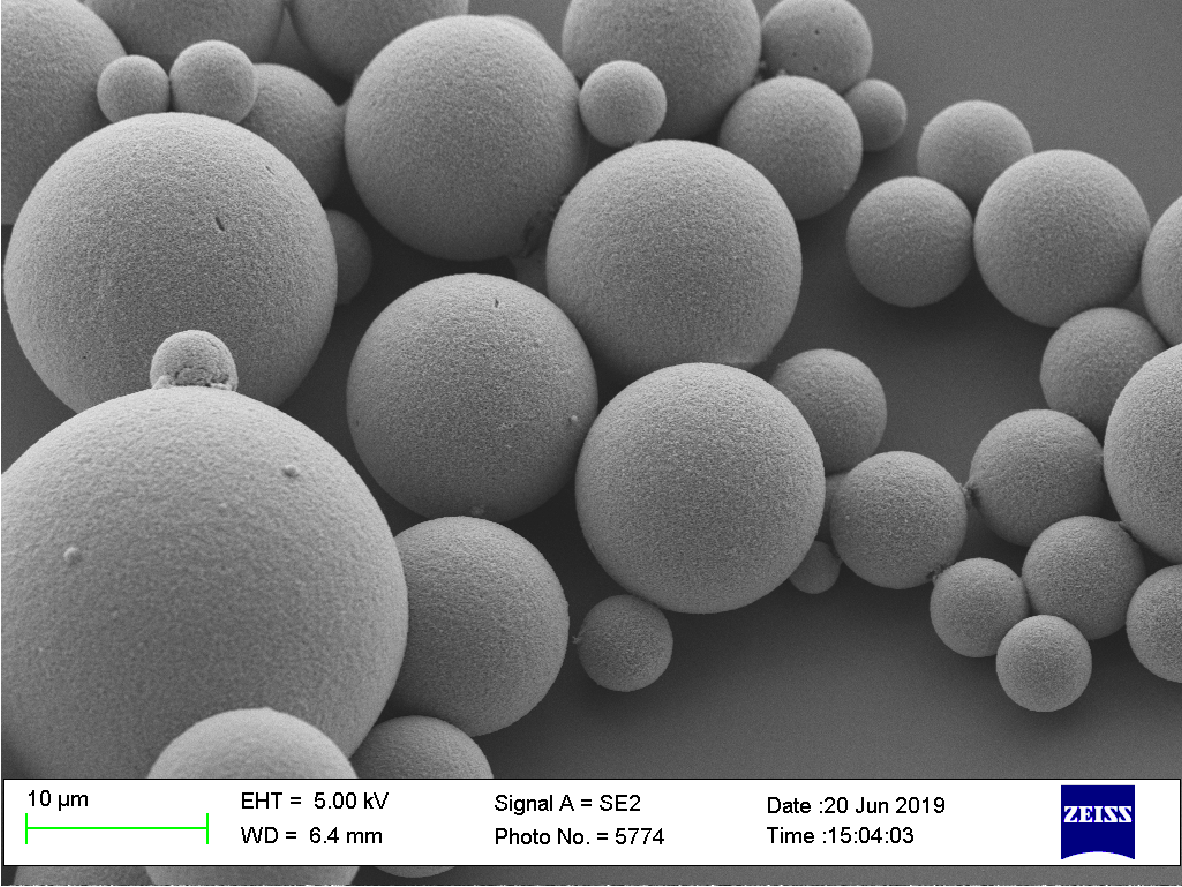} 
    \caption{Scanning electron micro-graph (SEM) of a group of assembled BaTiO$_3$ micro-spheres with different diameters. } \label{fig:BTOballs_SeveralSizes}%
\end{figure}

The assembling procedure developed here returns BaTiO$_3$ micro-spheres with diameters between 0.5-20~$\upmu$m. The quality of spherical shape is excellent over the entire size range, as can be seen from the SEM image in Fig.~\ref{fig:BTOballs_SeveralSizes}. In this particular case, the micro-spheres are loosely assembled in a cluster that allows us to image different sizes simultaneously. In this work, we investigate isolated micro-spheres, which can be found on the sample thanks to a grid of micron-size markers realized on the glass substrate by UV lithography and chromium deposition. 
%
\section{Measurement of the filling fraction}
\label{sec:ff_measurement}
\begin{figure}[h!]
    \centering
    \includegraphics[scale=1]{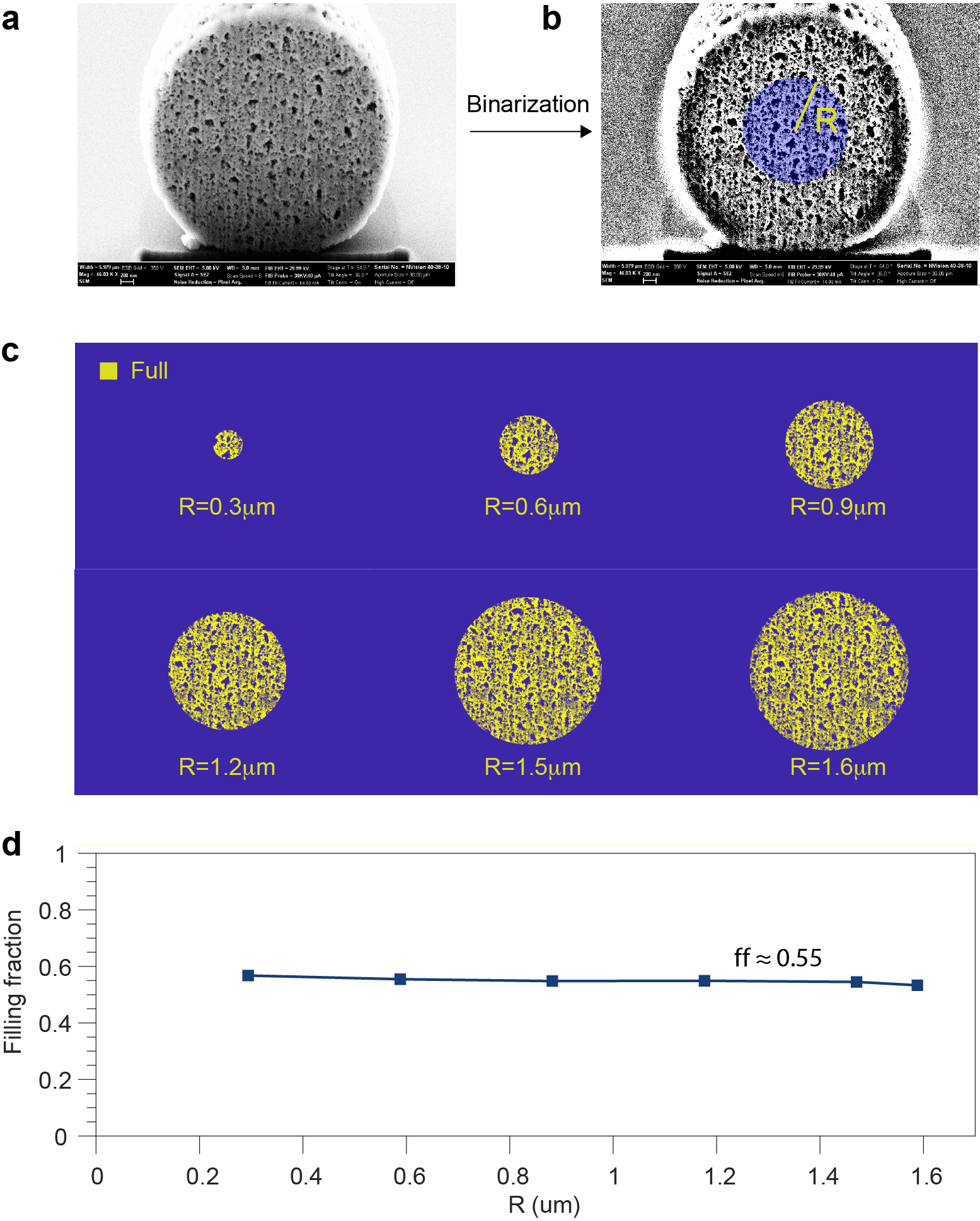} 
    \caption{Procedure to measure the filling fraction of an assembled BaTiO$_3$ micro-sphere. \textbf{a}, SEM image of the cross section of a micro-sphere realized by FIB. \textbf{b}, Binarized version of (\textbf{a}) obtained through the adaptive Otzu algorithm, where a circular region of radius $R$ is selected in the center. \textbf{c}, Selected binarized regions with increasing radii used to measure the filling fraction. The full part is in yellow, the empty part in blue. \textbf{d}, Measured filling fraction as a function of the radius of the selection region.}
	\label{fillingfraction}%
\end{figure}

The filling fraction of the assembled BaTiO$_3$ micro-spheres has been measured by image analysis of their  cross-sections obtained by focus ion beam (FIB) milling. Images are acquired with a scanning electron microscope (SEM) in-situ by applying  a tilt angle of 45$^\circ$ to the sample. The procedure is  performed with  a FIB-SEM NVision40-Zeiss. To minimize local charging effects, the micro-sphere is previously covered  with a 3~nm layer of platinum in a sputtering deposition chamber Elios600i. The sample is mounted on a planetary stage to ensure the uniformity of the platinum layer over the entire micro-sphere. The cross-section for a micro-sphere of about 3~$\upmu$m is shown in Fig.~\ref{fillingfraction}. The image is  tilt-corrected in the vertical direction by the acquisition software to measure exact lengths .

The cross-section image is binarized through the adaptive Otzu algorithm~\cite{otsu1979threshold}, implemented with the Matlab function \texttt{imbinirize}, which returns 0-value pixels for empty space and 1-value pixel for full (nano-crystals filled) space, see Fig.~\ref{fillingfraction}b. A circular region of interest (ROI) is selected by applying a circular mask of radius $R$ to the binarized image, as shown in Fig.~\ref{fillingfraction}b-c. The filling fraction $ff$ is measured on the ROI as $ff=N_{\textrm{full}}/N_{\textrm{TOT}}$, with $N_{\textrm{full}}$ the number of full pixels with value 1 and $N_{\textrm{TOT}}$  the total number of pixels. The radius of the ROI is gradually increased to test if the micro-sphere is homogeneously assembled. The considered ROI and the corresponding filling fraction are shown in Fig.~\ref{fillingfraction}c-d. We find $ff\approx0.55$ independently of the ROI size, showing a good homogeneity of the assembly. 

The same procedure has been applied to 5 different micro-spheres with radii between 1.2-5~$\upmu$m, providing comparable results. The measured $ff$ is consistent with the typical filling fraction of dried porous media realized by nano-particles assembly, which ranges between 50-70$\%$~\cite{zurcher2018evaporation}.  
\clearpage
%
%
\section{Linear microscopy and spectroscopy setup}
\label{sec:Linear}
\begin{figure}[]
    \centering
    \includegraphics[width=13cm]{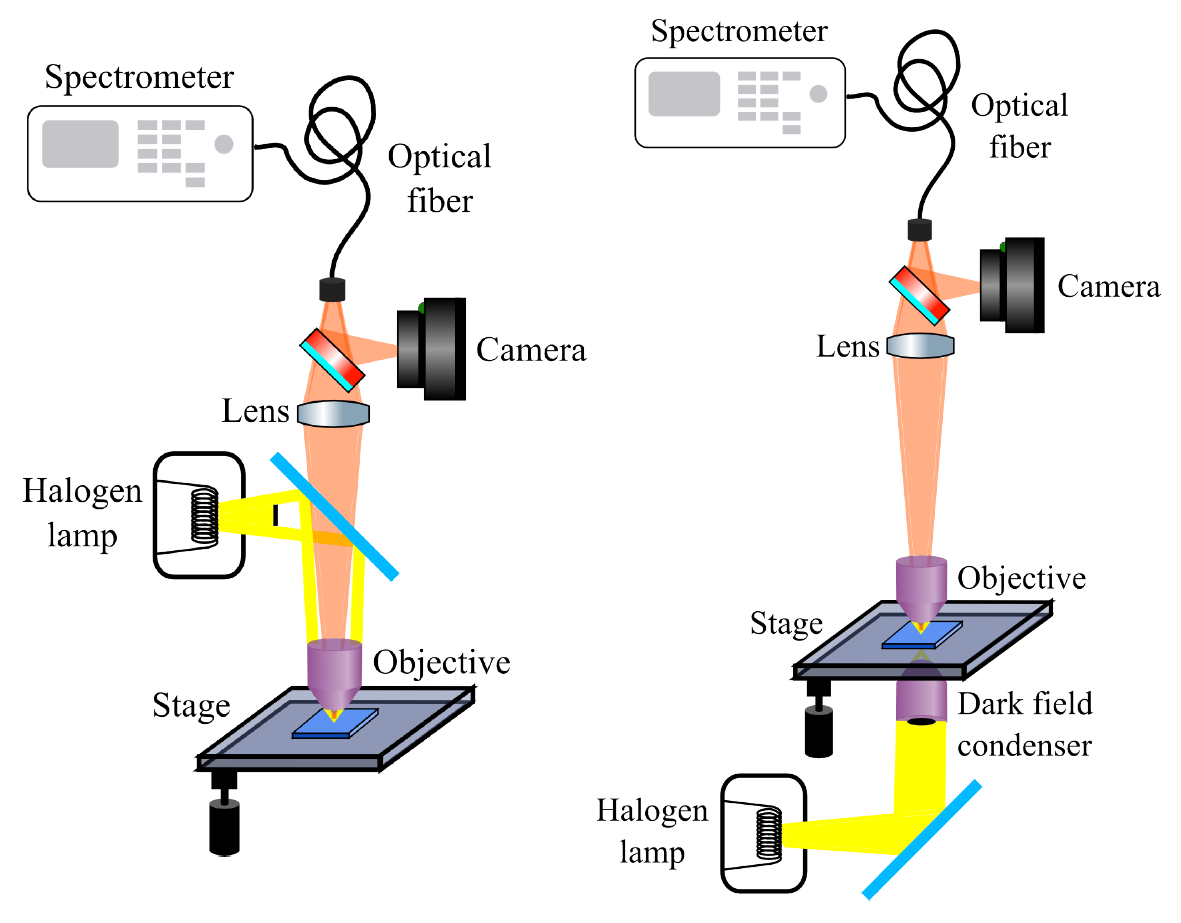} 
    \caption{Dark field spectroscopy setup in reflection (left) and transmission (right) configurations.}
	\label{darkfield}%
\end{figure}

The linear optical characterization of the micro-spheres has been performed with a Zeiss AxioImager.A1m  microscope customized for  spectral measurements. A sketch of the microscope setup is shown in Fig.~\ref{darkfield} for both the transmission and the reflection configuration. Samples are illuminated with a halogen lamp and light is collected with a 50X objective (Zeiss EC Epiplan Aphocromat, NA=0.95), providing a collection angle  $\theta=\arcsin(NA)=72^\circ$. Collected light is divided by a beam splitter and imaged with a lens on a CMOS camera to visualize the area of interest, while the other portion of light is imaged on the entrance of a multimode optical fiber (core size 200 $\upmu$m, NA=0.12). The fiber acts as a pinhole and collects only light coming from a small area in the center of the image with a field-of-view (FOV) of 4~$\upmu$m. Light is guided to a high-precision spectrometer (Acton SP-2356 from
Princeton Instruments, equipped with a thermoelectrically cooled CCD camera, Pixis 256E) for spectral analysis. The microscope works both in bright-field (BF) and dark-field (DF) configuration depending on the type of illumination selected. With the BF illumination light reaches the sample  homogeneously (Köhler illumination) and all transmitted/reflected light falling within the collection angle is acquired. With the DF illumination the central part of the illumination beam is blocked by a patch stop, which leaves only a ring of the illumination reaching the sample as a  focused empty cone. If the illumination is focused with an NA bigger than the NA of the collection element, illumination light falls out of the collection angle and only the scattered light is imaged. This technique allows one to obtain high optical contrast of quasi-transparent  and of very small objects.

DF microscopy in reflection is enabled by the specific design of the objective, which has two independent optical paths for the illumination and the collection. Typically, a high NA increases the chances to collect stray light. We minimized this effect by sticking a neutral density filter ND=8 below the glass substrate, which was index-matched with glycerine. DF imaging in transmission is possible when the NA of the condenser is larger than the NA of the collection objective. Since our condenser has an NA=0.8-0.9,  we had to reduce the NA of the objective to 0.7 by placing a rubber o-ring on the its back aperture. Even if approximated, the correction worked perfectly and high quality dark-field image could be obtained in transmission. We stress the importance of using an apochromat objective for this type of measurement to avoid chromatic aberration, which can affect spectral measurements by making them dependent on the sample-objective distance.     

Spectra were normalized as follows:
\begin{equation}
    \centering
    S(\lambda)=\frac{S(\lambda)_{\textrm{meas}}-S(\lambda)_{\textrm{BG}}}{S(\lambda)_{\textrm{Illum}}-S(\lambda)_{\textrm{DC}}} 
    \label{Eq:spectra_norm}
\end{equation}
where $S(\lambda)_{\textrm{meas}}$ is the scattering from the micro-sphere, $S(\lambda)_{\textrm{BG}}$ is the background (measured as the light scattered in an empty area near the micro-sphere), $S(\lambda)_{\textrm{Illum}}$ is the illumination and $S(\lambda)_{\textrm{DC}}$ are the dark counts. In the reflection configuration, $S(\lambda)_{\textrm{Illum}}$ was measured as the spectrum of the light reflected by a thick Teflon diffuser (99$\%$ reflectivity). In the transmission configuration, $S(\lambda)_{\textrm{Illum}}$ was  measured as the spectrum of the background in a completely dark environment.
%
%
\section{Effective-Medium-Mie (EMM) model}
\label{sec:EMM}
Assembled micro-spheres are modelled as homogeneous spheres with an effective refractive index $n_\mathrm{eff}$. This effective-medium approximation is then  combined with Mie theory to derive the wavelength dependence of two quantities: 1) the scattering cross section in the experimental configuration, used to describe the Dark-Field spectra of the micro-spheres; 2) the total internal energy, used within the random quasi-phase matching (RQPM) model for resonant structures (see Sec.~\ref{sec:RQPMmodel_resonant}),  to describe the SHG in the micro-spheres.  
In general, effective-medium mixing formulas are expected to work well for diluted composite media with a low index-contrast~\cite{zhuromskyy2017applicability}, which is not the case for the BaTiO$_3$ micro-spheres. However, when the size of the composite medium gets smaller (e.g. nanoparticles clusters, meso-crystals), the main requirement is to have  individual nanoparticles with an optical response dominated by the electric dipole term~\cite{voshchinnikov2007effective,zhuromskyy2017applicability, markel2016introduction}.
In our case, BaTiO$_3$ nano-crystals have a full dipole behaviour, with a size parameter $x\approx0.3$ (Rayleigh scattering regime), and the effective-medium approximation is a viable solution.
The advantage of using an effective-medium theory lies in its simplicity and in the possibility to describe the optical behaviour of a complex medium with a single parameter. Normally, the approximation gets weaker when light scattering from the constituent particles/enclosures is not negligible. In such a case an exact solution of Maxwell's equations is required~\cite{vettenburg2019calculating}.
\subsection{Effective refractive index with the Maxwell-Garnett mixing rule}
\begin{figure}[]
    \centering
    \includegraphics[scale=1]{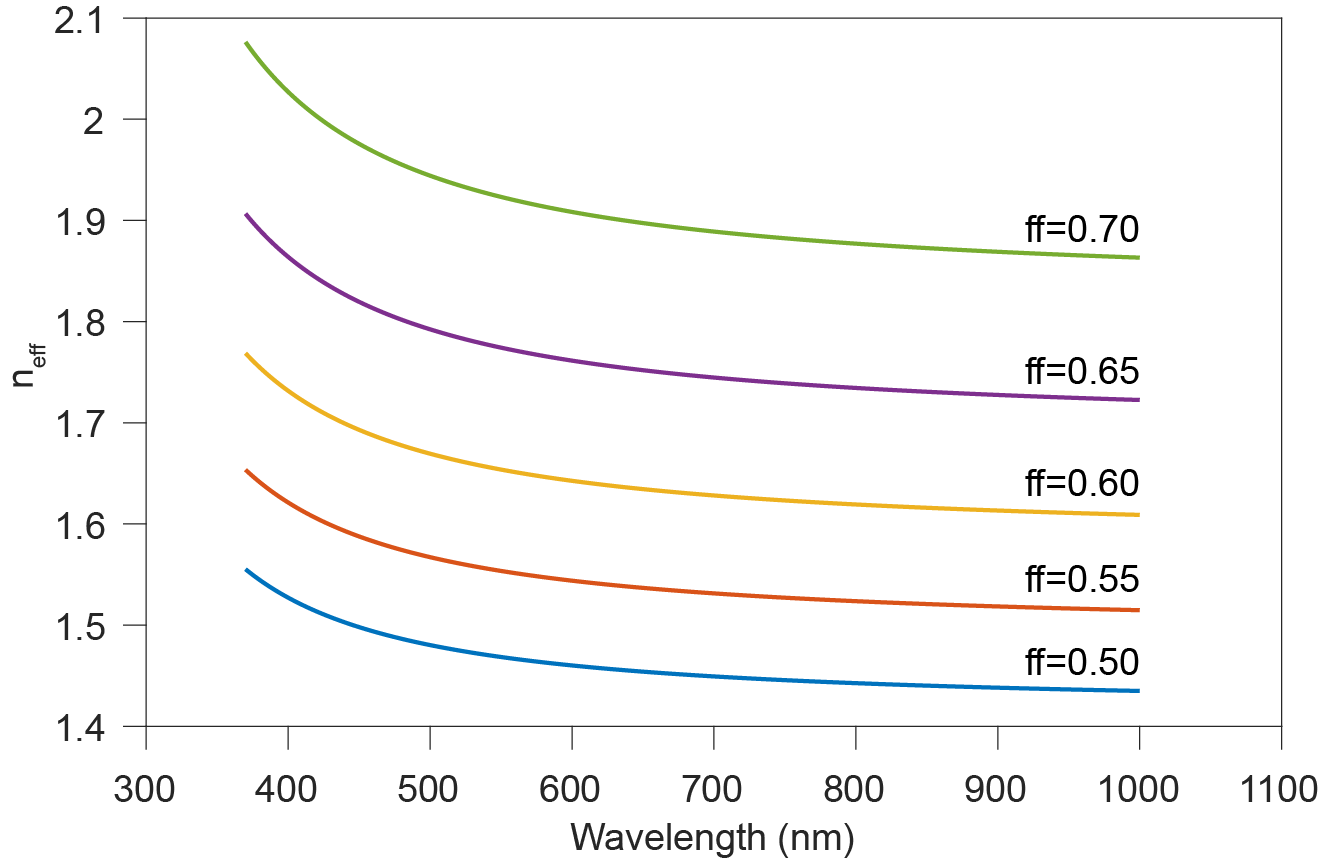} 
    \caption{Wavelength-dependent effective refractive indices for the BaTiO$_3$-air composite calculated with the Maxwell-Garnett mixing rule at different filling fractions. Assembled micro-spheres have a refractive index corresponding to $ff$=0.55. }
	\label{fig:neff_MG}%
\end{figure}

We calculated the effective refractive index of the assembled nanocrystals-air composite with the Maxwell-Garnett (MG) mixing rule reported in Eq.~\ref{eq:Maxwell-Garnett}, with $n_\mathrm{BTO}$ the average refractive index of bulk BaTiO$_3$ (see Sec.~\ref{sec:thematerial} and Fig.~\ref{fig:n_BTO}) and $ff$ the nano-crystals filling fractions of the composite. The wavelength-dependent effective refractive indices calculated at different filling fractions are shown in Fig.~\ref{fig:neff_MG}.
\begin{equation}
    n^2_\mathrm{eff}(\lambda)=\dfrac{1+\dfrac{1+2ff}{3}(n^2_\mathrm{BTO}(\lambda)-1)}{1+\dfrac{1-ff}{3}(n^2_\mathrm{BTO}(\lambda)-1)}
    \label{eq:Maxwell-Garnett}
\end{equation}
We employed the MG model, and not other effective-medium mixing rules as the Brugmann's rule, for the following reasons: firstly, recent works show that MG provides a better estimation of the effective refractive index for small and packed systems~\cite{zhuromskyy2017applicability}. Secondly, the values of the filling-fraction returned from the fit of the experimental data, performed by using the EMM model based on the MG formula, are in very good agreement with the filling fractions measured experimentally (see Sec.~\ref{sec:ff_measurement}).
\subsection{Expression of the micro-sphere's scattering cross-section and internal energy}
We employed the analytical Mie theory~\cite{bohren2008absorption} for spherical particles to calculate both the scattering cross-section and the internal energy of the micro-spheres in the experimental configuration. The cross-section is derived from the angle-dependent Mie scattering amplitudes $S_1$  and $S_2$  whose expressions are reported in Eq.~\ref{eq:Mie_sca_cros_sec1}-\ref{eq:Mie_sca_cros_sec2}, where $a_n,b_n$ are the Mie coefficients, $\pi_n$ and $\tau_n$ are combinations of Legendre functions defined in Bohren Huffman \cite{bohren2008absorption} p.94,  $x=ka$ is the size parameter with $a$ the radius of the sphere and $k =\frac{2\pi}{\lambda}$ the wave number in the ambient medium. 
S$_1$ and S$_2$ are the amplitudes of the scattered far-field component respectively orthogonal and parallel to the scattering plane.
All coefficients depend on the refractive index of the sphere, which here is the effective refractive index $n_\textrm{eff}$ of the micro-sphere. We do not report the full expressions of the Mie coefficient since they can be found in Bohren and Huffman~\cite{bohren2008absorption} at page 100. To compute Eq.~\ref{eq:Mie_sca_cros_sec1}-\ref{eq:Mie_sca_cros_sec2} we have implemented (in Matlab) the function \texttt{Mie\_S12} released by C. Mätzler~\cite{matzler2002matlab}.
\begin{equation}
    S_1(cos\theta)=\sum_{n=1}^\infty \frac{2n+1}{n(n+1)}(a_n\pi_n+b_n \tau_n)
    \label{eq:Mie_sca_cros_sec1}
\end{equation}   
    
\begin{equation}
    S_2(cos\theta)=\sum_{n=1}^\infty \frac{2n+1}{n(n+1)}(a_n\tau_n+b_n \pi_n)
    \label{eq:Mie_sca_cros_sec2}
\end{equation}
 To match the collection conditions of the dark-field scattering experiment, the scattering amplitudes $S_1$ and $S_2$ have been integrated over the collection angle $[108, 180^\circ]$ for the reflection configuration and over $[0, 50^\circ]$ for the transmission configuration. The unpolarized scattering intensity is calculated as $\mathrm{\sigma_{unpol}}=(S_1S_1^*+S_2S_2^*)/2$.  
 
To calculate the internal energy of the micro-spheres we have used the expression of the radial energy density in Eq.~\ref{eq:Mie_int_energy)}, which we have implemented (in Matlab) by using the function \texttt{Mie\_Esquare} released by C. Mätzler~\cite{matzler2002matlab}, and which we have integrated  over the entire volume of the micro-sphere. The coefficients  $c_n, d_n$ in Eq.~\ref{eq:Mie_int_energy)}  are the Mie coefficient and $m_n, n_n$ are combinations of spherical Bessel functions integrated over the angles $\Theta$ and $\phi$
\begin{equation}
   \biggl\langle|\mathbf{E}(r)|^2 \biggl\rangle=\dfrac{1}{4}\sum_{n=1}^\infty(m_n|c_n|^2+n_n|d_n|^2)
    \label{eq:Mie_int_energy)}
\end{equation}
\subsection{Effect of the substrate}   
\begin{figure}[t!]
    \centering
    \includegraphics[scale=1]{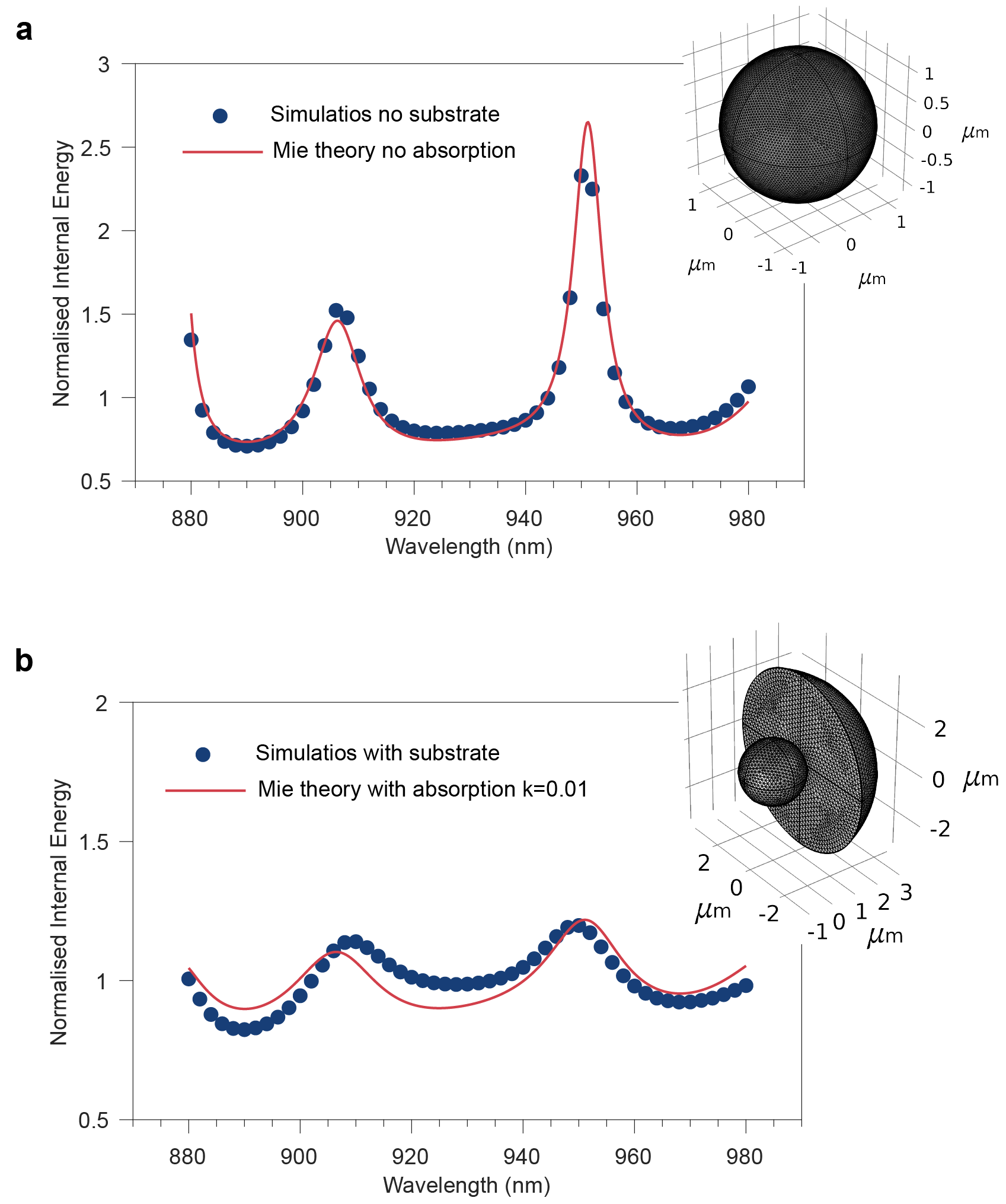}
    \caption{ Internal energy computed by FEM simulations and by analytical Mie theory for a sphere of radius $r=1200$~nm and $n_\textrm{eff}=1.6$, for a configuration \textbf{a}, without substrate and\textbf{b} with the sphere on top of a glass substrate. The insets in the figures show the configurations used in the simulations.}
	\label{fig:effect_substrate}%
\end{figure}

Analytical Mie theory does not consider the presence of a substrate near the particle, which is expected to affect its scattering behaviour. We investigated this effect by finite-element-method (FEM) simulations (in Comsol) of light scattering from a sphere of radius 1.2 $\upmu$m and $n_\textrm{eff}=1.6$ on a 1~mm thick glass substrate of refractive index 1.3 . We run the simulations for a plane-wave illumination at several wavelengths in the range 880-980~nm and calculated the total scattering cross-section and the internal energy of the sphere. For comparison we run simulations with the same configuration without the substrate. Results are shown in Fig.~\ref{fig:effect_substrate} for the internal energy, but apply similarly for the scattering cross-section. Peaks of the energy spectrum are dumped and widened due to energy out-coupling  into the substrate. In other words, the substrate introduces losses and the internal energy at the resonances is reduced. We show in Fig.~\ref{fig:effect_substrate} that the analytical Mie model can take these losses into account by adding a complex part $ik$ to the refractive index with an absorption coefficient typically of $k$=0.005. The agreement between theory and simulation is very good and provides the possibility to use an analytical theory to describe the resonant optical behaviour of the real micro-spheres, avoiding time-consuming simulations. As we see in Fig.~\ref{fig:effect_substrate}b the substrate also introduces a slight shift in the resonant peaks that the Mie theory cannot describe. 
%
%
\section{Nonlinear microscopy and spectroscopy setup}
\label{sec:nonlinear_setup}
\begin{figure}[]
    \centering
    \includegraphics[scale=1]{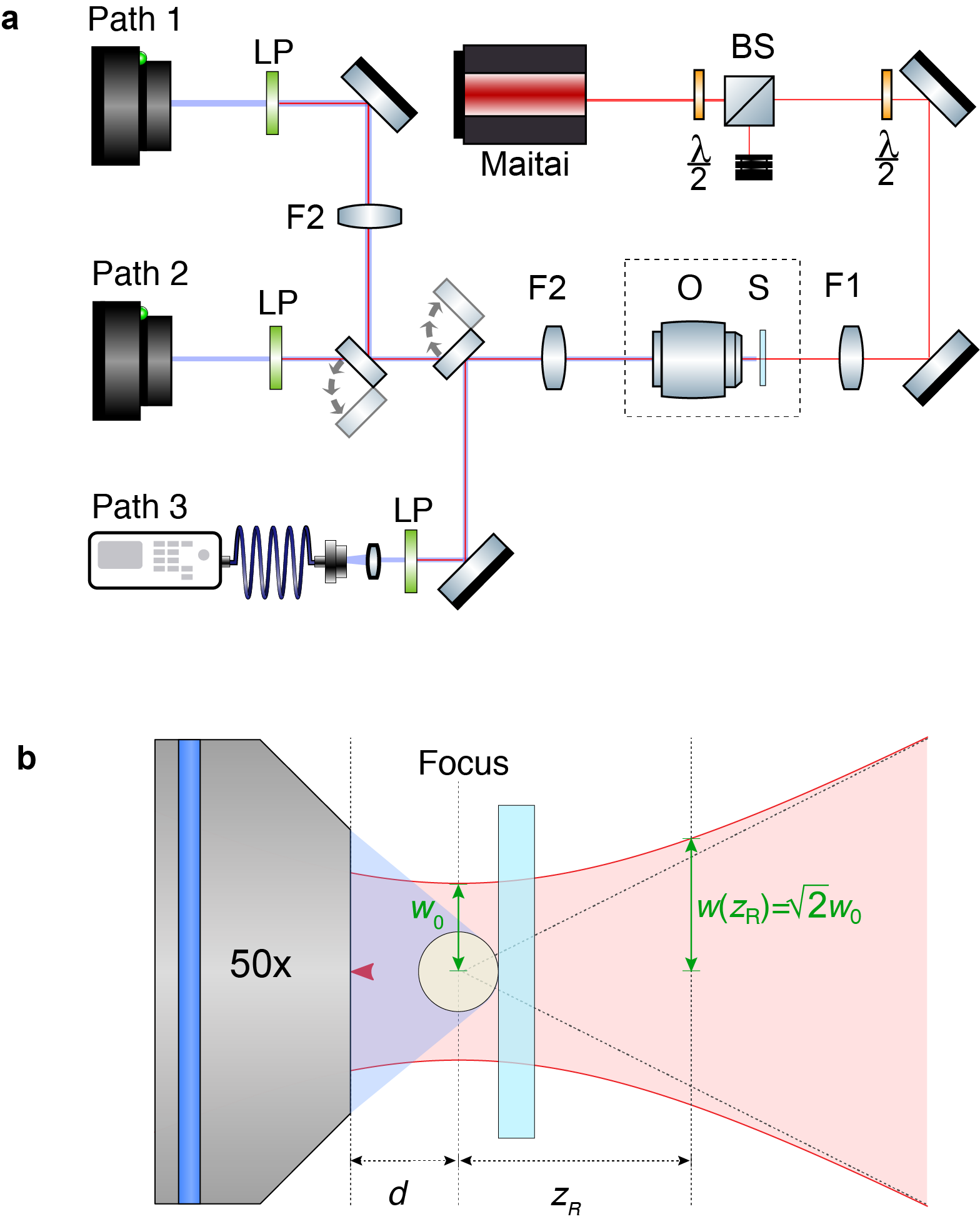} 
    \caption{Schematic of the setup used for the  nonlinear optical characterization. \textbf{a}, Beam paths to excite and collect the SHG from micro-spheres. ($\frac{\lambda}{2}$) Wave plate (BS) Beam splitter (F1,F2) Lens (S) Sample (O) Objective (LP) Low pass filter. \textbf{b}, Schematic of the focusing configuration on the sample. The fundamental beam passes from right to left through the micro-sphere (yellow). The emitted SHG (blue) is collected by the objective. ($w_0$) Beam waist (d) working distance ($z_R$) Rayleigh length.}
	\label{fig:setup}%
\end{figure}
A scheme of the setup employed to perform the nonlinear optical characterization of the micro-spheres is shown in Fig.~\ref{fig:setup}a.  Ultra-fast laser pulses (120~fs) at the fundamental frequency $\omega$ (pump) are generated by a  Titanium-Sapphire laser (Spectra-Physics MaiTai HP) with a repetition rate of 80~MHz, an average output power of 2.5 W and central wavelength tunable in the range 690-1040~nm. To adjust the power of the beam, a half-wave-plate ($\frac{\lambda}{2}$) is used to rotate the beam polarization, followed by a Glan-Taylor polarizing beam-splitter (BS) allowing only one polarization angle to pass through. A second half-wave-plate sets the linear polarization of the beam before it hits the sample. This second wave-plate is rotated with a motorized stage controlled by the computer. To have a uniform illumination of the micro-sphere, we loosely focused the pump beam with a plano-convex lens (F1) with a focal length of 75~mm, corresponding to a beam waist of 20 $\upmu$m.
The sample is oriented such that the beam passes first through the glass before hitting the micro-sphere. A magnified illustration of the sample position is shown in Fig.~\ref{fig:setup}b.
The collection objective (Zeiss 50x EC Epiplan apochromat, $NA=0.95$, working distance $d=0.5$~mm) is placed after the sample to collect the SHG from the micro-sphere at $2\omega$. The objective is mounted on a high-precision translation stage to allow for accurate focusing. We always selected the center of the micro-sphere as the focal plane. Thanks to the large numerical aperture, we collect most of the forward-generated light with a collection angle $\theta=\arcsin(NA)=72^\circ$. 
Collected light can go through 3 different paths. In path 1 we image the back-focal-plane (BFP) of the objective with 2 plano-convex lenses (F2, focal length 150 mm) on the camera (Andor Zyla 4.24. In path 2 we image the sample on the camera with the same lens F2. 
In path 3 collected light is redirected with a flipping mirror and focused into a fiber that guides the light to a spectrometer (Andor Shamrock 303i-B with camera Newton 920). In all the paths light passes through a low-pass filter (LP) before reaching the camera/detector to remove the fundamental beam and leave only the SHG. The SHG emitted at the focal plane is in focus on the camera, while the rest is collected and measured as out-of-focus signal. The total SHG is indeed independent from the position of the focal plane within the micro-sphere.
Images of the rear plane of the micro-spheres at the pump wavelength ($\omega$) were obtained with the same setup by removing the low-pass filter. Since the SHG signal is 5 to 7 orders of magnitudes weaker than the pump, it was not necessary to filter out the SHG.

\subsection{Wavelength calibration}
The setup has been calibrated for operating in the range 880-980~nm by normalizing the experimental data by a curve that considers the wavelength dependence of the quantum efficiency of the camera, of the transmission through the low-pass filters (BG39) and through the objective. The full normalization formula is reported in Eq.~\ref{eq:SHG_spect_norm}. The focal position is not significantly affected by the change of the pump wavelength. The pulse width and the repetition rate are constant, therefore they do not alter the peak power in the explored range of wavelengths.
\begin{equation}
    \mathrm{SHG_{corrected}}=\dfrac{\mathrm{SHG_{measured}}}{2 \cdot \mathrm{T(\lambda)_{BG39}} \cdot  T(\lambda)_{\mathrm{OBJ}} \cdot \mathrm{QE(\lambda)_{camera}}}
    \label{eq:SHG_spect_norm}
\end{equation}
%
\subsection{Imaging of the back-focal-plane (BFP) of the objective}
\begin{figure}[ht]
    \centering
    \includegraphics[scale=1]{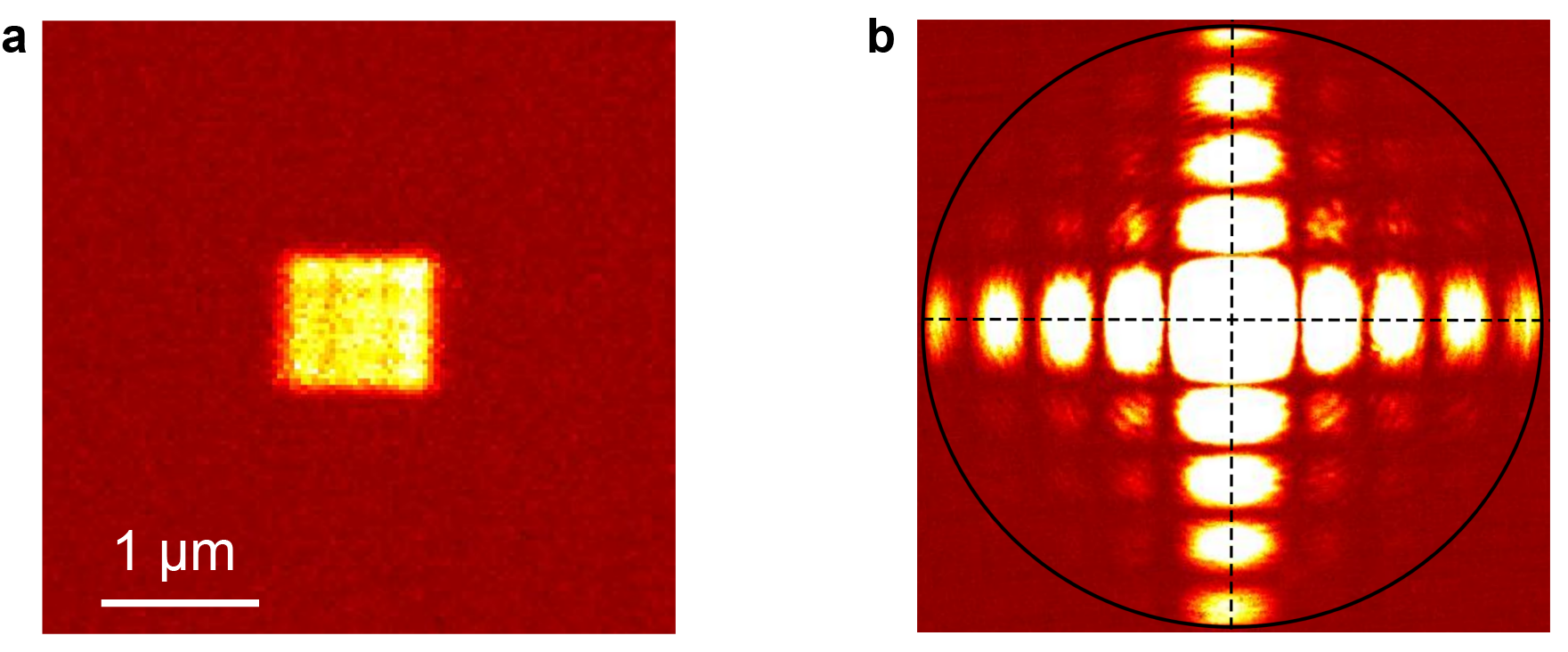} 
    \caption{\textbf{a}, Real image and \textbf{b}, back-focal-plane image of a square aperture of 1~$\upmu$m in the chromium layer deposited over a glass slide. The circle represents the maximal numerical aperture of the Zeiss apochromat objective, corresponding to 72$^\circ$}.
	\label{fig:BFP}
\end{figure}

The alignment and the distances along the BFP path of the set-up have been checked by using a reference sample made out of a chromium-covered glass slide with squared holes (from one to a few microns) in the chromium layer. An optical image of the reference sample is shown in Fig.~\ref{fig:BFP}a.

The Fourier distribution expected by a squared aperture of size \textit{a} is given by
\begin{equation}
    \mathrm{I(x,y)} \sim \sinc^2 \left( \dfrac{\pi a}{\lambda f}x \right) \sinc^2 \left( \dfrac{\pi a}{\lambda f}y \right) 
    \label{eq: Back focal plane}
\end{equation}
where $\lambda$ is the wavelength of light, $f$ is the focal distance of the lens and $a$ is the size of the square.
The BFP of the objective is in focus on the camera when the pattern shown in Fig.~\ref{fig:BFP}b appears, which matches the one predicted by Eq.~\ref{eq: Back focal plane}. Once the BFP was in focus, without moving any optics, the test sample was substituted by the sample with the micro-spheres.

\newpage
%
\section{Spectrum and power dependence of the measured SHG}
\begin{figure}[ht!]
    \centering
    \includegraphics[scale=1]{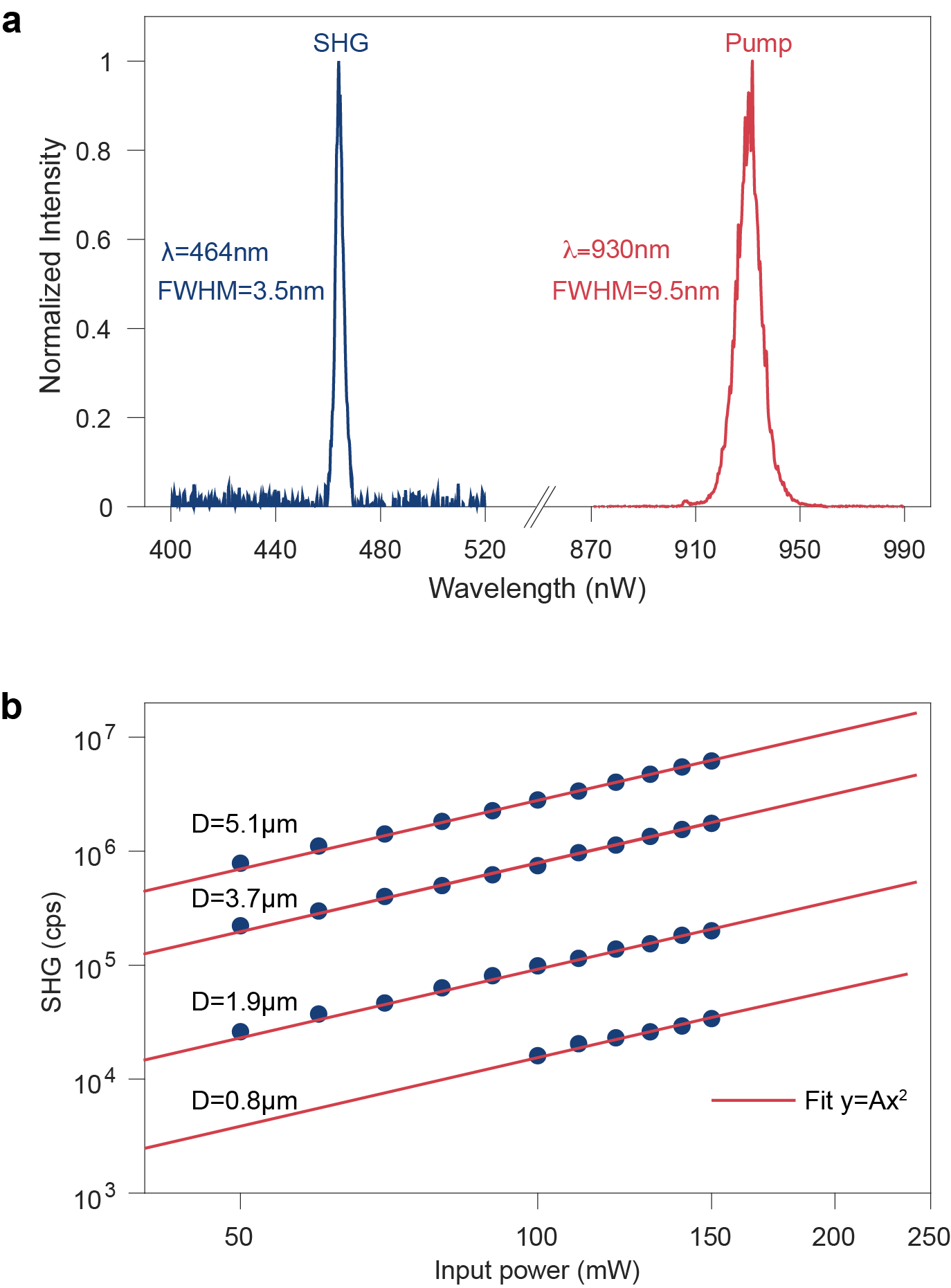} 
    \caption{Characterization of the SHG from assembled BaTiO$_3$ microspheres. \textbf{a}, Spectrum of the SHG compared to the pump spectrum. \textbf{b}, Quadratic dependence of the  SHG on the pump power for micro-spheres of different sizes. Axis are in log-scale. Power values refer to the entire beam. Slopes indicates the quadratic dependencies expected from SHG while vertical shifts show that the SHG efficiency grows with the size of the micro-sphere. The efficiency of the smallest micro-sphere is too low to observe SHG below 100~mW.}
	\label{fig:Spectrum_Power}%
\end{figure}

Identification of the SHG has been first carried out by spectral characterization of the light passing through the low-pass filter, which has been coupled into an optical fiber and guided to a spectrometer, as shown in Fig.~\ref{fig:setup}a. A spectral characterization of the pump has been performed in the same configuration by removing the low-pass filter. Measured spectra are reported in Fig.~\ref{fig:Spectrum_Power}, together with the peak wavelengths and the full-width-half-maximum (FWHM) bandwidths. The signal wavelength is half the pump wavelength. The signal bandwidth is also about half the pump bandwidth. Both observations  confirm the presence of SHG. As a secondary check, we tested the dependence of the measured signal on the pump power, which is expected to be purely quadratic for SHG. Measurements are shown in Fig.~\ref{fig:setup}b for several sizes of the micro-sphere. Clear quadratic dependencies are observed for all sizes, further confirming the presence of SHG. For a fixed pump power, the SHG continually increase with the size of the  micro-sphere, as expected in the random quasi-phase-matching regime. 

\section{Simulations of SHG in BaTiO$_3$ disordered assemblies}
We performed simulations of disordered BaTiO$_3$ assemblies, using an approach similar to previously established simulations for isotropic materials\cite{chen2019non,vidal2006generation}, extended to be applicable to a birefringent material, as BaTiO$_3$. The model considers a 1D stick of crystalline domains. Each domain has a variable thickness and crystal orientation (length $X_n$ and orientation $O_n$ for the $n^{th}$ domain). The ordinary (o) and extraordinary (e) parts of the fundamental wave are propagated separately through each domain, generating SH-waves at the end of every domain, given by the electric field along the ordinary and the extraordinary direction ($u = o/e$):

\begin{equation}
    E_{\text{generated}}^{u}(2\omega, X_n)  = \sum_{vw}\frac{i(2\omega) ^2 }{2\epsilon_0 c^2 k^u_3} \cdot \big(\sum_i\hat{e}^u_i\cdot 2\epsilon_0\sum_{jk}d_{ijk}E_j^vE_k^w\big) \bigg(\frac{e^{i\Delta k^{u,vw}X_n}-1}{i\Delta k^{u,vw}}\bigg)e^{ik^u_3X_n}
    \label{eq:SHGfield_DisorderedStick_Birefringent}
\end{equation}

with 
\begin{itemize}[label={}]
    \setlength\itemsep{-0.5em}
    \item $u,v,w \in \{o,e\}$, specifying the polarization of the beam component
    \item $\omega$, the fundamental frequency
    \item $X_n$, the size of the n$^{th}$ domain
    \item $\epsilon_0 = 8.854187817\cdot10^{-12}$ F/m, the vacuum permittivity
    \item $c = 2.99792458\cdot 10^8$ m/s, the speed of light in a vacuum
    \item $k^u_3$, the $\vec{k}$-vector of the second harmonic along $u$
    \item $\hat{e}^u$, the unit vector along the $o/e$-direction
    \item $d_{ijk}$, the second order non-linear tensor. Its contracted matrix form is specified in equation \ref{eq:BTO_tensor}
    \item $E^v$ and $E^w$, the electric field of the fundamental along $v$ and $w$
    \item $\Delta k^{u,vw}$, the phasemismatch as specified in equation \ref{eq:phase-mismatch}
\end{itemize}

Those SH-waves are then propagated in a similar fashion throughout the entire stick and summed up to the end (for computational efficiency the summation is done after every domain, propagating only the resulting waves). For this propagation and to use the second harmonic tensor $\chi^{(2)}$, the fields must be transformed into the reference frame of the respective domain, and decomposed into the $o/e$ direction. These transformation were done using an Euler transformation.

\subsection{Efficiency comparison with a bulk crystal}
\begin{figure}[ht!]
    \centering
    \includegraphics[scale=1]{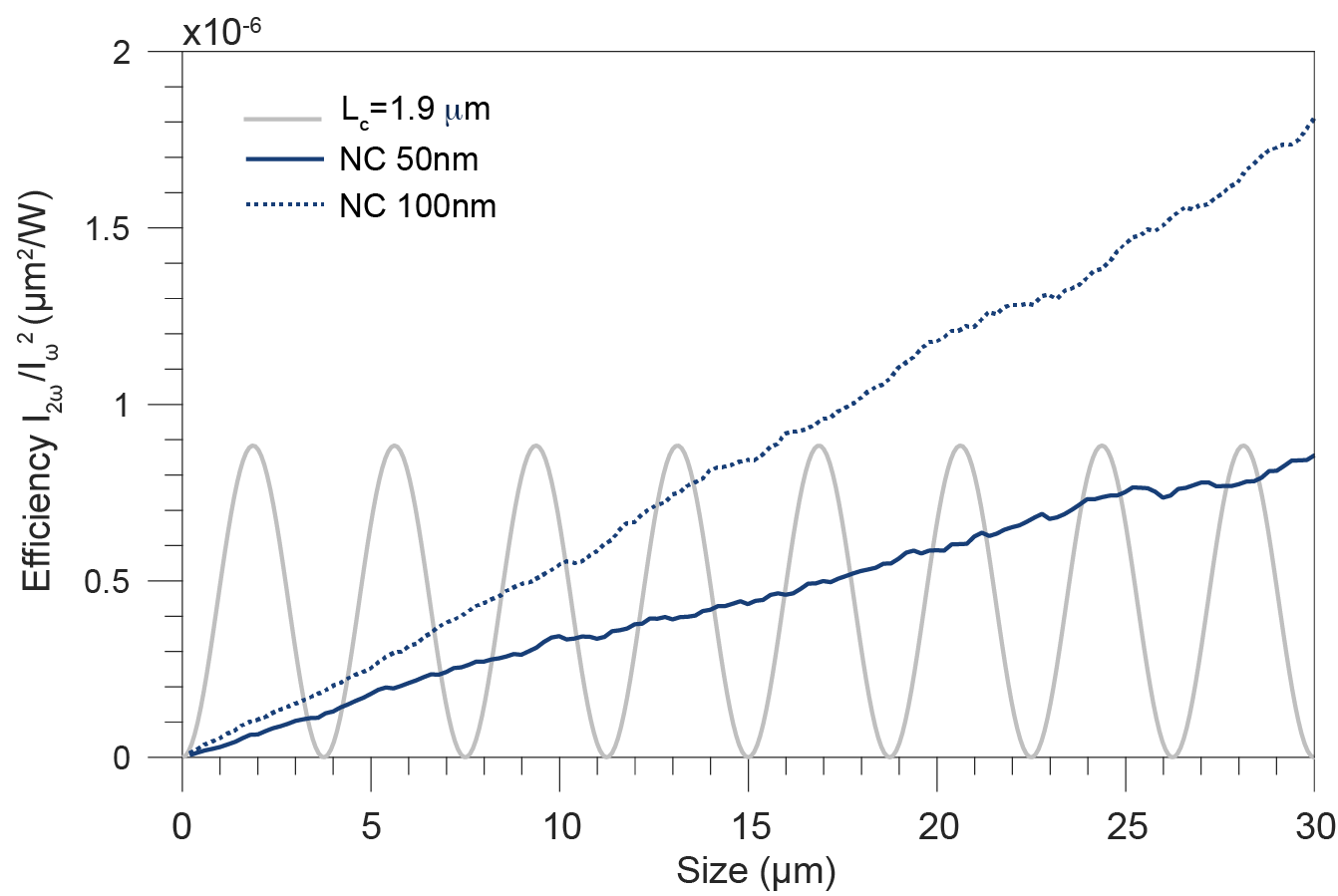}
    \caption{Calculated SHG efficiencies for a cubic BaTiO$_3$ single crystal (gray line) in its optimal orientation (longest coherence length $L_c$) and for two cubic random assemblies with 50~nm (blue solid line) and with 100~nm (blue dotted line) nano-crystal (NC) domains, at 930~nm. This represents the efficiency of the system for an input intensity of 1 W/$\upmu$m$^2$. }
    \label{fig:comparison_with_bulk_crystal}
\end{figure}

By relying on Eq.~\ref{eq:SHGfield_DisorderedStick_Birefringent} we computed the SHG efficiency $I_{2\omega}$/$I_{\omega}^2$ of disordered BaTiO$_3$ cubes by considering disordered sticks with $N$  domains and by assembling $N\times N$ sticks in the transverse dimensions. Then, we computed the SHG efficiency in a BaTiO$_3$ crystal cube of the same size. The resulting efficiencies of the bulk crystal and of the assembly (for two domain sizes) are shown in Fig.~\ref{fig:comparison_with_bulk_crystal}.
The efficiency of the single crystals in its “optimal” orientation (gray line) periodically drops to zero, whenever the length of the crystal along the propagation direction is equal to an even number of coherence lengths. 
Results shown in Fig.~\ref{fig:comparison_with_bulk_crystal} are the same shown in the inset of Fig~3p of the main manuscript. Since here we consider the intensity of the systems (Power/Surface) the transverse size of the system, i.e. perpendicular to the propagation direction, is irrelevant and we obtain a linear scaling of the RQPM with the length of the assembly. The efficiency of the disordered system  with a mean domain size of 50~nm (blue solid line)  is comparable to that of the bulk crystal in the considered size range and surpasses the ideally oriented crystal at around 30~$\upmu m$. The efficiency of the disordered system when using a 100~nm domain size (blue dotted line) beats the single crystal already at around 15~$\upmu m$. One can see that doubling the domain size leads to a doubling of the SHG intensity (solid to dotted blue line). The reason for this is, that with a constant system size $S$, an increase in the domain size $s_{\text{domain}}$ will in return reduce the number of domains $N$ within the system, with a cubic scaling with domain size: $I \propto N \propto (1/s_{\text{domain}})$. At the same time the increase in the domain size $s_{\text{domain}}$ along the propagation direction leads to a quadratic increase of intensity within each crystal $I \propto s_{\text{domain}}^2$. In total, this leads to a linear dependence of the intensity with the domain size $s_{\text{domain}}$ in a 3D system of constant size: $I \propto s_{\text{domain}}^2/s_{\text{domain}} \propto s_{\text{domain}}$. 

\section{Random walk model for the random quasi-phase-matching (RQPM)\\ in resonant structures}
\label{sec:RQPMmodel_resonant}
\begin{figure}[t]
    \centering
    \includegraphics[scale=1.05]{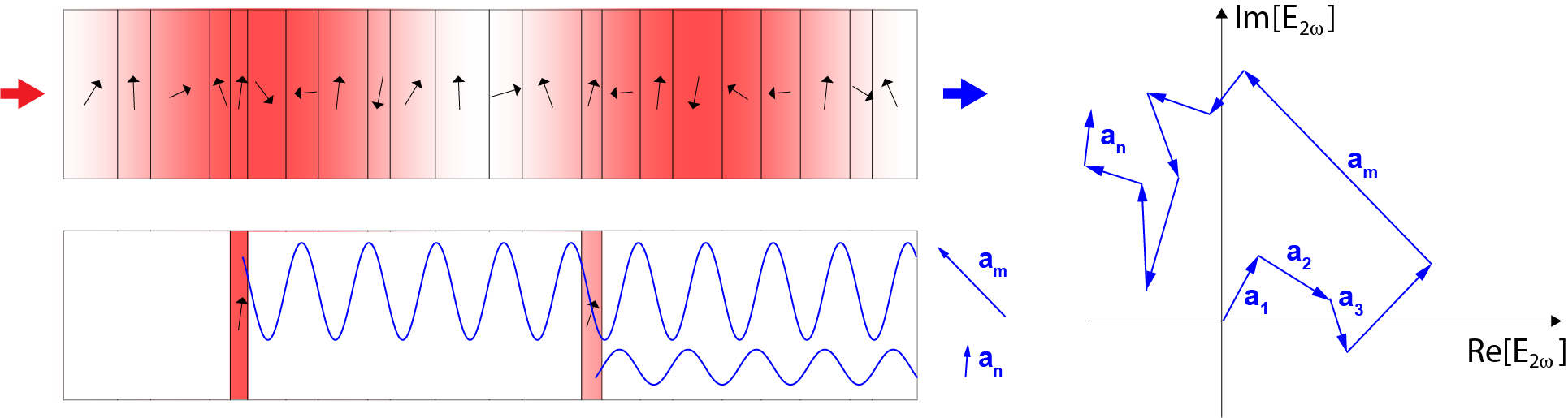}
    \caption{Sketch of a one-dimensional $\chi^{(2)}$-disordered system  and  effect of an optical mode on the RQPM mechanism. The color modulation within the disordered crystalline stick indicates the intensity distribution of the pump, which here is a $\sin^2{(x)}$ as an example. Crystalline domains in a region of high field-enhancement (red) generate a SH-wave with a larger amplitude compared to the others. The corresponding phasors ($a_m$) are longer steps of the random walk in the SH complex plane, which enhance the RQPM generation by increasing the mean square displacement. }
    \label{fig:resonantRQPM}
\end{figure}

We consider a 1D disordered array of $N$ poly-dispersed crystalline domains with random orientations, as sketched in Fig.~\ref{fig:resonantRQPM}.  The SHG field at the end of the array is given by the sum of $N$ phasors, whose phase and amplitude depends on the disorder configuration, as expressed in Eq.~\ref{eq:SHGfield_DisorderedStick_Synthetic}. This expression is a synthetic version of Eq.~\ref{eq:SHGfield_DisorderedStick_Birefringent}, in which we consider the phase terms randomly distributed in $[0,2\pi]$ and in which we highlight the dependence of the amplitudes on the randomly distributed lengths $L_i$ and orientations $s_i$ of the domains. 
\begin{equation}
    E_\mathrm{SHG}=\sum_{i}^N A(L_i,s_i) e^{i\phi_i}
    \label{eq:SHGfield_DisorderedStick_Synthetic}
\end{equation}
Equation~\ref{eq:SHGfield_DisorderedStick_Synthetic} describes a random walk in the complex plane of the SHG field. Accordingly, the SHG intensity is given by the mean square displacement (MSD) of this random walk  $ I_{SHG}=\langle(\sum_{i}^N A(L_i,s_i)\cdot e^{i\phi_i})^2\rangle$, where $\langle...\rangle$ indicates the ensemble averaging. 
By assuming that all domains are equally illuminated by the pump beam and that there is no correlations between the contributions of the phasors, the MSD grows linearly with the number of steps, such that $I_{SHG}\propto N$, without any particular wavelength-dependent modulation. This is the well known result of RQPM~\cite{baudrier2004random,vidal2006generation,chen2019non}.

Now, we assume that the disordered array confines light and  sustains optical modes, both at $\omega$ and $2\omega$. In such a case the specific patter of the modes defines an inhomogeneous distribution of the pump  and of the SHG field, see Fig.~\ref{fig:resonantRQPM}. We assume that the spatial features of the modes evolve on a scale larger than the mean size of the domains (large-scale modes), such that the fields can be  considered constant over the single domain. This situation can be described by a domain-dependent SHG enhancement factor $\xi_i(\omega, 2\omega)=\omega^2 F_i^2(\omega)F_i(2\omega)$ that is multiplied to the amplitude $A_i$ and that enhances or decreases the SHG field of each domain, as in an homogeneous nonlinear resonator~\cite{zhang2014enhanced}. $F_i(\omega)$ and $F_i(2\omega)$ are the field enhancement of the pump and of the SHG respectively. Equation~\ref{eq:SHGfield_DisorderedStick_Synthetic} modifies to Eq.~\ref{eq:SHGfield_DisorderedStick_Enhanced}, corresponding to the MSD of a random walk with a mode-dependent step-lengths distribution, in which some steps  contribute more than others if the corresponding domain is in a high-enhancement region     
\begin{equation}
    I_{SHG}\propto\langle(\sum_{i}^N \xi_i(\omega, 2\omega) A(L_i,s_i)\cdot e^{i\phi_i})^2\rangle.
    \label{eq:SHGfield_DisorderedStick_Enhanced}
\end{equation}
We consider the standard situation of RQPM, where the system is disordered in the $\chi^{(2)}$ spatial distribution, but it is homogeneous in the $\chi^{(1)}$ spatial distribution, i.e. the refractive index. This way, the modes are determined by the geometry and are independent from specific disorder configurations. Within this assumption, we run random-walk simulations for arbitrary field distributions of the modes (with finite mean and variance) and found that after ensemble averaging the summation over $\xi_i(\omega, 2\omega)$ in Eq.~\ref{eq:SHGfield_DisorderedStick_Enhanced} can be taken out of the summation, as expressed in Eq.~\ref{eq:SHGfield_DisorderedStick_Enhanced_2}

\begin{equation}
    I_{SHG}\propto(\sum_{i}^N \xi_i(\omega, 2\omega))^2 \cdot \langle( \sum_{i}^N  A(L_i,s_i)\cdot e^{i\phi_i})^2\rangle 
    \label{eq:SHGfield_DisorderedStick_Enhanced_2}
\end{equation}
We evaluate the total contribution of the field enhancement as
\begin{equation}
  \sum_{i}^N \xi_i(\omega, 2\omega) = \omega^2 \sum_{i}^N  F_i^2(\omega)F_i(2\omega) \\
  \propto \omega^2 E_\textrm{int}(\omega) \sqrt{E_\textrm{int}(2\omega)}
  \label{eq:SHGfield_DisorderedStick_Enhanced_3}
\end{equation}
where $E_\textrm{int}=\sum_{i}^NF_i^2 $ is the internal energy of the system at the considered wavelength and where we neglect the contribution of the spatial-overlap integral of the pump and of the SH mode~\cite{lin2016cavity}. The final expression for the SHG intensity is reported in Eq.~\ref{eq:SHGfield_DisorderedStick_Enhanced_4} 
\begin{equation}
    I_\mathrm{SHG} \propto \omega^2 E_\textrm{int}^2(\omega) E_\textrm{int}(2\omega) N.
    \label{eq:SHGfield_DisorderedStick_Enhanced_4}
\end{equation}
We find the expected linear dependence on the number of domains and an additional modulation term $\omega^2 E_\textrm{int}^2(\omega) E_\textrm{int}(2\omega)$ accounting for the resonant enhancement, in which the leading factor depends on the internal energy at the pump wavelength. In the case of Mie resonances within a sphere, the wavelength-dependent internal energy can be calculated by Eq.~\ref{eq:Mie_int_energy)}. For our micro-spheres, this calculation is exact for the pump since we have a plane-wave illumination, but it is an approximation for the SHG, since real sources are randomly placed withing the sphere.  
\subsection{Fitting procedure}

We have fitted the data of the wavelength-dependent SHG from the micro-spheres with Eq.~\ref{eq:SHGfield_DisorderedStick_Enhanced_4} in the wavelength range 880-980~nm, by normalizing both measurements and theory to their mean values. The internal energies  have been calculated in the EMM approximation. The free parameters in the least-squared fitting procedure were the filling fraction ($ff$) and the absorption coefficient ($k$).
Best-fit values are reported in Tab.~\ref{tab:ffandk}. We can see that the estimated filling fractions  are around 0.55, in good agreement with results from the image analysis in Sec.~\ref{sec:ff_measurement} and from the linear spectroscopy in Sec.~\ref{sec:Linear}. Also, the estimated $k$, which are all around 0.01, are compatible with the expected losses introduced by the substrate, calculated in Sec. \ref{sec:EMM}.
We stress that the parameters obtained by this fitting procedure rely on independent data since they correspond to different samples, supporting the validity of our analytical model and showing a remarkable consistency of the work.

\begin{table}
\centering
\begin{tabular}{c|c|c}
\toprule
\emph{Radius} [nm] & \emph{Filling fraction} & \emph{Absorption (k)} \\   
\midrule 
609 & 0.506 & 0.003 \\ 
\hline 
975 & 0.462 & 0.013 \\ 
\hline 
1005 & 0.598 & 0.012 \\ 
\hline 
1018 & 0.590 & 0.019 \\ 
\hline 
1195 & 0.554 & 0.012 \\ 
\hline 
1261 & 0.652 & 0.014 \\ 
\hline 
1495 & 0.458 & 0.008 \\ 
\hline 
2383 & 0.570 & 0.007 \\ 
\hline 
4319 & 0.546 & 0.005 \\ 
\bottomrule
\end{tabular} 
\caption{Radius of the assemblies measured by SEM, filling fraction and absorption coefficient $k$ (for a refractive index $n+ik$) extracted from the fitting procedure.}
\label{tab:ffandk}
\end{table}

\newpage

\bibliography{supplementaries.bib}